\RequirePackage{fix-cm}
\documentclass[twocolumn,epjc3]{svjour3}  
\smartqed  
\RequirePackage{graphicx}
\journalname{Eur. Phys. J. C}
\usepackage{amssymb}
\usepackage[fleqn]{amsmath}
\usepackage[english]{babel}

\newcommand{\ee}{${\rm e}^+{\rm e}^-$}
\newcommand{\ppbar}{${\rm p}{\rm p}\hskip-7pt\hbox{$^{^{(\!-\!)}}$}$}

\begin{document}

\title{A Monte-Carlo generator for statistical hadronization in 
high energy \ee collisions
}


\author{C. Bignamini\thanksref{e1,addr1,addr2}, 
	     F. Becattini\thanksref{e2,addr3,addr4} \and 
	     F. Piccinini\thanksref{e3,addr2}}

\thankstext{e1}{e-mail: christopher.bignamini@pv.infn.it}
\thankstext{e2}{e-mail: becattini@fi.infn.it}
\thankstext{e3}{e-mail: fulvio.piccinini@pv.infn.it}

%
%
\institute{Dipartimento di Fisica, Universit\`a di Pavia - Pavia, Italy \label{addr1} \and
                INFN Sezione di Pavia - Pavia, Italy \label{addr2} \and
	       Dipartimento di Fisica e Astronomia, Universit\`a di Firenze - Firenze, Italy \label{addr3} \and
                INFN Sezione di Firenze - Firenze, Italy \label{addr4}}

\date{Received: date / Accepted: date}

\maketitle

\begin{abstract}
We present a Monte-Carlo implementation of the Statistical Hadronization Model 
in \ee collisions. The physical scheme is based on the statistical hadronization
of massive clusters produced by the event generator \texttt{Herwig} within
the microcanonical ensemble. We present a preliminary comparison of several observables
with measurements in \ee collisions at the Z peak. Although a fine tuning of the 
model parameters is not carried out, a general good agreement between its predictions
and data is found. 

\end{abstract}

\section{Introduction}
\label{intro}

The Statistical Hadronization Model (SHM) has proved to be successful in reproducing 
essential features of the hadronization process, such as the multiplicities and the transverse 
momentum spectra of many hadronic species in high energy elementary and nuclear collisions 
(see e.g. \cite{becareview} and references therein). Recently, it has been shown that 
this model is also able to satisfactorily reproduce the rates of multi-hadronic 
exclusive channels in \ee annihilation at low energy \cite{BecFerr}. Accor\-ding to the 
SHM, hadronization proceeds through the formation of extended massive colorless 
objects called {\em clusters} or {\em fireballs} emitting hadrons according to a pure statistical 
law, i.e. all multihadronic states compatible with conservation laws are equally likely. 
The number and properties of these clusters are determined by the pre-hadronization dynamical process 
and cannot be predicted within the SHM. Therefore, in \ee and \ppbar collisions at high 
e\-nergy \mbox{($\sqrt s >$ 10 GeV)}, the SHM calculations have been carried out with supplementary 
assumptions, in order to obtain analytical or semi-analy\-tical formulae to be compared directly with
the data. Such is the case of the so-called {\em Equivalent Global Cluster} picture 
\cite{becareview,BecPass}, in which a further assumption is invoked concerning the distribution 
of charges and masses among the hadroni\-zing pre-hadronic clusters; this allows to
make calculations in the more manageable canonical ensemble.

While these supplementary assumptions have been very useful to point out the statistical
features of hadronization, it would be desirable to perform tests of the SHM independently
thereof, at the most fundamental level of formulation. As clusters are generally small, 
calculations in a basic SHM framework involve the computation of averages in the microcanonical
ensemble, i.e. with exact conservation of energy and momentum, besides that of abelian 
charges (baryon number, strangeness, electric charge, charm and beauty). This kind of calculation
has been recently carried out (with the further complication of angular momentum 
conservation and more) for \ee annihilation at low e\-nergy \cite{BecFerr}. Yet, in
this case, only one cluster was involved instead of many as it is generally the case for
high energy collisions. 

In a multi-cluster environment, the only possible way to cope with the problem of statistical 
hadronization is a Monte-Carlo integration. In this paper a prototype of Monte-Carlo event generator 
for the SHM is presented and the results obtained by testing it on \ee collisions at $\sqrt s =91.2$ 
GeV center of mass energy, i.e. at the Z peak, are discussed. 

For the present study, the developed Monte-Carlo hadronization code has been interfaced to the 
\texttt{Herwig6510} event generator \cite{HerwigOld}, which is used to simulate the \ee col\-lisions 
except for the hadronization phase. This step is instead performed using the SHM 
hadronization module. This particular choice for the external generator is owing to the similarity 
between the input clusters required by the SHM and those produced by \texttt{Herwig6510} \cite{ClusterH}.   
In our framework these clusters are used as starting point for the statistical hadronization process. 
Finally, the unstable hadron decays are performed with the \texttt{Herwig6510} generator itself.

The paper is organized as follows: in Sec. \ref{sec:SingleClu} the general framework of the SHM is 
summarized and the algorithms for the simulation of single cluster hadronization are presented. 
In Sec. \ref{sec:StdCluMod} a comparison between the theoretical formulation of the SHM and those 
of the Cluster Model implemented in other Monte Carlo generators, such as \texttt{Herwig6510}, is reported. 
The computational methods for multi-cluster hadronization in the SHM framework are then discussed 
in Sec.~\ref{sec:MultiClusterAlgo}. In Sec. \ref{sec:SHMPreliminaryTest} the full event simulation setup, 
involving the external event generator, is described, as well as results obtained for a preliminary adjustment 
of free parameters. The impact on SHM predictions of variations of some \texttt{Herwig6510} free parameters 
involved in the cluster formation procedure is also discussed. In Sec. \ref{sec:NumericalResults} the 
results obtained for \ee collisions at the Z peak with the SHM-based hadronization module are presented 
and compared with the corresponding \texttt{Herwig6510} predictions and LEP experimental data. 
Conclusions and future developments are outlined in Sec. \ref{sec:Concl}.

\section{Statistical hadronization of single clusters}
\label{sec:SingleClu}

As it has been mentioned in the Introduction, the cornerstones of the SHM are:
\begin{itemize}
\item in a high energy collision, as a result of the pre-hadronic dynamical process, 
a set of extended massive colorless objects called clusters or fireballs are 
formed;
\item all multi-hadronic states confined within the cluster and compatible with 
conservation laws are equally likely.
\end{itemize}

The second postulate gives rise to a definite mathematical formula \cite{becareview,BecFerr} 
involving (pseudo)projector operators associated to conservation laws, for 
the probability of observing a specific final state from a single cluster decay. 
In principle, conservation laws to be implemented include energy-momentum, angular momentum, 
parity and all of the internal symmetries relevant to strong interactions. However, like 
all other Monte-Carlo generators, we confine ourselves to abelian charges (electric charge, 
strangeness, baryonic number, charm and beauty), in addition to energy-momentum. In fact, 
in a multi-cluster environment, relevant to this work, dealing with non-abelian 
symmetry groups is overwhelmingly difficult if clusters take their origin from a pure quantum 
state and are entangled with each other at the hadronization stage. 

In the simpler scheme of additive abelian charges and energy-momentum conservation, the 
probability of a multi-hadronic final state for a single cluster can be derived as described 
in \cite{BecFerr1,BecFerr2}. Particularly, neglecting quantum statistics and quantum field
effects, the microcanonical weight $\Omega_{\{N_{j}\}}$ for the multi-hadronic channel 
$\{N_{j}\}$ made of $N_{1}$, $N_{2}$,..., $N_{K}$  hadrons of species $1,\ldots,K$ is given 
by \cite{BecFerr1}:
\begin{equation}
\begin{split}
\Omega_{\{N_{j}\}} &= \prod_{j=1}^{K}\frac{1}{N_{j}!}\left[\frac{\left(2J_{j}+1\right)V}
{\left(2\pi\right)^{3}}\right]^{N_{j}} \times \\
& \displaystyle\int\prod_{i=1}^{N} d^{3}p_{i}\delta^{4}(P - P_{f})
\delta_{\mathbf{Q},\mathbf{Q}_{f}},
\end{split}
\label{eq:microweigth}
\end{equation} 
with $P_{f}= \sum_{i=1}^{N}p_{i}$ and $\mathbf{Q}_{f}=\sum_{i=1}^{N}\mathbf{Q}_{i}$, where 
$p_{i}$ is the 4-momentum of the \mbox{$i$-th} hadron in the channel while 
$\mathbf{Q}_{i} = (B_i,Q_i,S_i,\ldots)$ is the array collecting its abelian charge values 
(baryon number, charge, strangeness, etc.), and $N$ is the total number of particles in the 
channel. $P$, $\mathbf{Q}$ and $V$ are the cluster 
\mbox{4-momentum}, abelian charges and volume respectively. Finally, it must be noted that a sum 
over particle spin states has been performed: the $J_{j}$ quantities in the corresponding contributions 
to the total weight are the spin modulus of the single hadron species.
The microcanonical probability $p(\{N_{j}\})$ of the considered channel is given by:
\begin{equation}
p(\{N_{j}\}) = \frac{\Omega_{\{N_{j}\}}}{\Omega},
\label{eq:microprob}
\end{equation} 
where $\Omega$ is defined as the sum over all possible channels
\begin{equation}
\Omega = \displaystyle\sum_{\{N_{j}\}}\Omega_{\{N_{j}\}},
\label{eq:micropart}
\end{equation}
i.e. the microcanonical partition function. Finally, the microcanonical
probability density in phase space can be written as: 
\begin{equation}
\begin{split}
p\left(\{N_{j}\},\{p_{j}\}\right) &= \frac{\delta^{4}(P - P_{f})\delta_{\mathbf{Q},
\mathbf{Q}_{f}}}{\Omega}\times\\
&\displaystyle\prod_{j=1}^{K}\frac{1}{N_{j}!}\left[\frac{\left(2J_{j}+1\right)V}{\left(2\pi\right)^{3}}\right]^{N_{j}},
\end{split}
\label{eq:microtrans}
\end{equation} 
where $\{p_{j}\} = p_{1}, p_{2},..., p_{N}$ is the set of four-momenta of the
particles in the channel $\{N_{j}\}$.

The analysis of particle multiplicities has shown \cite{becareview} that particles carrying
strange valence quarks need an additional suppression with respect to the pure statistical
predictions. This has led to the introduction of the \emph{strangeness suppression parameter} 
$\gamma_{\textnormal{S}}$, which, in order to fit its definition for inclusive multiplicities,
has to multiply $\Omega_{\{N_{j}\}}$ of Eq. \ref{eq:microweigth} as follows:
\begin{equation}
\Omega_{\{N_{j}\}} \rightarrow \Omega_{\{N_{j}\}}\gamma_{\textnormal{S}}^{\sum_{j=1}^{K}N_{j}s_{j}},
\label{eq:strangesuppr1}
\end{equation}
where $N_{j}$ is the number of hadrons of species $j$ in the channel $\{N_{j}\}$ and $s_{j}$ 
is the total number of valence strange and antistrange quarks contained in the $j$-th hadron 
species. Extra strangeness suppression also applies to light unflavored mesons, like the 
$\eta$ meson, which are actually a flavorless superposition of states with a wave function 
of the general form:
\begin{equation}
C_{u}u\bar{u}+C_{d}d\bar{d}+C_{s}s\bar{s},
\label{eq:superposition1}
\end{equation}
with $|C_{u}|^{2}+|C_{d}|^{2}+|C_{s}|^{2} = 1$. For this kind of hadrons only the $s\bar{s}$ 
part of the wave function is supposed to be suppressed, with a suppression factor given by
\begin{equation}
1- |C_{s}|^{2} + |C_{s}|^{2}\gamma_{\textnormal{S}}^{2}.
\label{eq:strangesuppr2}
\end{equation}
The $\gamma_{\textnormal{S}}$ parameter and the cluster energy density $\rho$ 
are the only free parameters of the model, in its microcanonical formulation. The cluster 
proper energy density $\rho$ plays the role of conversion factor between the cluster volume 
$V$, present in Eqs. \ref{eq:microweigth}-\ref{eq:microtrans}, and the cluster mass $M$ 
through the relationship
\begin{equation}
\rho = \frac{M}{V}.
\label{eq:endens}
\end{equation}
%

\subsection{Single cluster hadronization algorithm}
\label{subsec:SingleCluAlgo}

The simulation of the cluster microcanonical hadronization is mainly based on the algorithms 
described in \cite{BecFerr3} and here summarized. More in detail, the single cluster
hadronization process involves two main steps, namely the sampling of the hadronization channel 
and of the corresponding kinematical configuration.

Due to the large number of available decay channels for a hadronizing cluster, an efficient 
sampling algorithm is needed to obtain a fast and optimized hadronization simulation: for the 
present case the sampling function $\Pi$  has been defined, as discussed in \cite{BecFerr3}, 
as the multi-species multiplicity distribution of the SHM grandcanonical formulation
 
\begin{equation}
\Pi_{\{N_{j}\}}=\prod_{j=1}^{K}\exp\left(-\nu_{j}\right)\frac{\nu_{j}^{N_{j}}}{N_{j}!},
\label{eq:samplf1}
\end{equation} 
where $\nu_{j}$ is the mean number of particles of the $j$-th type, $K$ the number of included hadronic 
species and $N_{j}$ the number of particles of kind $j$ contained in the channel $\{N_{j}\}$. 
The mean multiplicities $\nu_{j}$ are free parameters which should be set with the aim of obtaining 
the most efficient sampling function. The abelian charge conservation is then introduced in the 
channel sampling procedure based on Eq. \ref{eq:samplf1} in the following way: the whole set of hadrons 
is divided into eleven sampling groups, namely, (anti)bottomed hadrons, (anti)charmed hadrons, light (anti)baryons, 
light strange (anti)mesons, light charged (anti)mesons with zero strangeness and neutral light mesons. 
Moreover, the following feature of the multi-poissonian function is used: considering for simplicity only 
the baryon, antibaryon and meson hadron groups and the relative charge conservations, the original 
sampling function with an extra Kronecker delta, added to impose the conservation of abelian charges, 
can be written as

\begin{equation}
\begin{split}
&\prod_{j=1}^{K}\exp\left(-\nu_{j}\right)\dfrac{\nu_{j}^{N_{j}}}{N_{j}!}\delta_{\sum_{i}N_{i}\mathbf{q}_{i},\mathbf{Q}} =
\prod_{j=1}^{K}\pi_{j}\left(N_{j}\right)\delta_{\sum_{i}N_{i}\mathbf{q}_{i},\mathbf{Q}} =\\ 
&\prod_{bar}\pi_{j}^{b}\left(N_{j}\right)\prod_{antibar}\pi_{j}^{\bar{b}}\left(N_{j}\right)\prod_{mes}\pi_{j}^{m}\left(N_{j}\right)\delta_{\sum_{i}N_{i}\mathbf{q}_{i},\mathbf{Q}}=\\    
&\Pi_{b}\left(N_{b}\right)\Pi_{\bar{b}}\left(N_{\bar{b}}\right)\times \\
&P(N_{1}^{b},N_{2}^{b},...\mid N_{b})P(N_{1}^{\bar{b}},N_{2}^{\bar{b}},...\mid N_{\bar{b}})\times\\
&\prod_{mes}\pi_{j}^{m}\left(N_{j}\right)\delta_{\sum_{i}N_{i}\mathbf{q}_{i},\mathbf{Q}},
\end{split}
\label{eq:sampltransf1}
\end{equation}
where

\begin{equation}
\Pi_{x}\left(N_{x}\right)=\exp\left(-\nu_{x}\right)\dfrac{\nu_{x}^{N_{x}}}{N_{x}!}
\label{eq:sampltranff2}
\end{equation} 
is the poissonian distribution of the total number of baryons ($x=b$) or antibaryons ($x=\bar{b}$) 
and where the corresponding mean multiplicity $\nu_{x}$ is given by the sum of the single baryon or 
antibaryon mean multiplicities. The functions $P$ are the conditioned multinomial 
distributions of the single hadronic species, namely

\begin{equation}
P\left(N_{1}^{x},N_{1}^{x},...\mid N_{x}\right)\propto N_{x}!\prod\dfrac{\nu_{j}^{N_{j}^{x}}}{N_{j}^{x}},
\label{eq:multinomial}
\end{equation} 
with, again, $x=b$ for baryons and $x=\bar{b}$ for antibaryons.

With the above decomposition of the original multipoissonian distribution, the following sampling 
algorithm can be used to perform the hadronization channel generation:
\begin{enumerate}
\item extract randomly the number of baryons and antibaryons according to the distributions 
$\Pi_{b}\left(N_{b}\right)$ and $\Pi_{\bar{b}}\left(N_{\bar{b}}\right)$;
\item check whether the baryonic number is conserved. If not reject the sampling and go to point 1, 
otherwise generate the single baryons and antibaryons using the multinomial distribution of Eq. \ref{eq:multinomial};
\item extract the single mesons using the initial multipoisson distributions;
\item check the conservation of the remaining abelian charges, taking into account the already 
generated baryons and antibaryons. If the check is not passed start again with point 1, otherwise the 
sampled channel is momentarily accepted and the corresponding phase space availability can be verified.
\end{enumerate}

The adoption of the described sampling procedure, instead of the independent sampling of the single
hadronic species, which would follow from Eq. \ref{eq:samplf1}, is motivated by the reduction 
of random number extractions which occur in case of rejection of the sampled channel due to the abelian 
charge conservation: the independent sampling procedure would require, for each channel sampling attempt, 
the extraction of $K$ random numbers, where $K$ is the number of hadron species included in the hadronization 
procedure, which would be lost in case of event rejection. The above algorithm, on the other hand, allows 
to stop and start again the sampling procedure, in case of charge conservation failure, when a number of 
random extractions smaller than $K$ has been performed by checking the conservation of the single abelian 
charges during the sampling. 

The only exception to the above sampling procedure is represented by the heavy flavored 
hadron sampling step, required when a heavy flavored cluster must be hadronized. For these objects, which are 
supposed to hadronize into channels containing one heavy hadron and a set of light particles, the heavy 
hadron is sampled with probability $P_{h}$ of the form 

\begin{equation}
P_{h}\propto e^{-a m_{h}},
\label{eq:heavysampl}
\end{equation}
with $a>0$ and where $m_{h}$ is the hadron mass. This particular choice for the heavy flavored hadron 
sampling function is justified by two reasons: on one side it allows to avoid the strong undersampling
which would follow from the usage of the standard sampling method (Eq. \ref{eq:samplf1}) also for these
hadronic species. On the other side, the above function allows to approximatively take into account the
heavy hadron relative production probabilities due to their mass differences. 
The remaining part of the channel is then randomly chosen using the same algorithm applied for the hadronization 
of light clusters. This modification in the sampling function for the heavy flavored clusters allows to obtain a more efficient 
channel generation algorithm with respect to the standard procedure previously described.
 
Given a hadronization channel composition, a kinematical configuration must be generated: in the present work, 
a sampling algorithm inspired by the multi-parti\-cle decay phase space integration described in \cite{PDG}\footnote
{The detailed description of the adopted phase space sampling algorithm can be found in \cite{phdthesis}.} 
is used. The starting point to obtain the kinematical configuration and the corresponding weight, for a decay 
channel of a cluster of mass $M$, containing $N$ particles with masses $m_{1},m_{2},...,m_{N}$, is the 
phase space integral of Eq. \ref{eq:microweigth}:

\begin{equation}
PS(M,m_{1},m_{2},...,m_{N}) = \displaystyle\int\prod_{i=1}^{N} d^{3}p_{i}\delta^{4}\left(P - \displaystyle\sum_{i=1}^{N}p_{i}\right).
\label{eq:kingen1}
\end{equation} 
As discussed in \cite{PDG}, the following condition holds for the relativistic $N$-body phase space element:

\begin{equation}
\begin{split}
\mathrm{d}\Phi_{N}(P;p_{1},p_{2},...,p_{N}) &= \mathrm{d}\Phi_{N-J+1}(P;Q,p_{J+1},...,p_{N}) \times \\ 
& \mathrm{d}\Phi_{J}(Q;p_{1},p_{2},...,p_{J})(2\pi)^{3}\mathrm{d}Q^{2}
\end{split}
\label{eq:kingen2}
\end{equation} 
for $2\leq J<N$, where

\begin{equation}
\mathrm{d}\Phi_{N}(P;p_{1},p_{2},...,p_{N}) = \delta^{4}\left(P - \displaystyle\sum_{i=1}^{N}p_{i}\right)\displaystyle\prod_{i=1}^{N}\displaystyle\dfrac{\mathrm{d}^{3}p_{i}}{(2\pi)^{3}2E_{i}}
\label{eq:kingen3}
\end{equation} 
and

\begin{equation*}
Q^{2}=\left(\displaystyle\sum_{i=1}^{J}E_{i}\right)^{2} - \left|\displaystyle\sum_{i=1}^{J}\mathbf{p}_{i}\right|^{2}. 
\end{equation*}
The non-relativistic phase space integral of Eq. \ref{eq:kingen1} can be rewritten making use of the relativistic 
phase space element of Eq. \ref{eq:kingen3} as follows:

\begin{equation*}
PS(M,m_{1},m_{2},...,m_{N}) = 
\end{equation*}
\begin{equation*}
\displaystyle\int\prod_{i=1}^{N} \dfrac{\mathrm{d}^{3}p_{i}}{(2\pi)^{3}2E_{i}}\delta^{4}\left(P - \displaystyle\sum_{i=1}^{N}p_{i}\right)\displaystyle\prod_{i=1}^{N}(2\pi)^{3}2E_{i}=
\end{equation*}
\begin{equation}
\displaystyle\int\displaystyle \mathrm{d}\Phi_{N}\left(P;p_{1},p_{2},...,p_{N}\right)\prod_{i=1}^{N}(2\pi)^{3}2E_{i}.
\label{eq:kingen4}
\end{equation} 
The integration/sampling procedure adopted here is based on the property of Eq. \ref{eq:kingen2}, iteratively 
applied in order to deal with 2-body decays only as described by

\begin{equation}
\begin{split}
&PS(M,m_{1},m_{2},...,m_{N}) = \\
&\displaystyle\int\displaystyle\prod_{i=0}^{N-3} \mathrm{d}\Phi_{2}\left(q_{i};p_{i+1},q_{i+1}\right)(2\pi)^{3}\mathrm{d}q_{i+1}^{2}\times\\
&\mathrm{d}\Phi_{2}\left(q_{N-2};p_{N-1},p_{N}\right)\displaystyle\prod_{i=1}^{N}(2\pi)^{3}2E_{i},
\end{split}
\label{eq:kingen5}
\end{equation}
with $q_{0}=(M,\mathbf{0})$ and where the integration limits of the $q_{i}^{2}$ virtualities are given by

\begin{equation}
\begin{split}
q_{i}^{2\ min} &= \left(\displaystyle\sum_{j=i+1}^{N}m_{j}\right)^{2}\\
q_{i}^{2\ max} &= \left(\sqrt{(q_{i-1})^{2}}-m_{i}\right)^{2}.
\end{split}
\label{eq:kingenlim}
\end{equation}
By solving the 2-body phase space integrals, Eq. \ref{eq:kingen5} becomes

\begin{equation}
\begin{split}
&PS(M,m_{1},m_{2},...,m_{N}) = \\
&\frac{1}{2^{N-2}}\displaystyle\prod_{i=1}^{N-2}\int\displaystyle\mathrm{d}q_{i}^{2}\int\displaystyle \mathrm{d}\Omega_{\mathbf{\bar{p}}_{i}}
\frac{|\mathbf{\bar{p}}_{i}|}{\bar{p}_{i}(0)+\bar{q}_{i}(0)}\times\\
&\int\displaystyle \mathrm{d}\Omega_{\mathbf{\bar{p}}_{N-1}}\frac{|\mathbf{\bar{p}}_{N-1}|}{\bar{p}_{N-1}(0)+\bar{p}_{N}(0)}\displaystyle\prod_{i=1}^{N}E_{i},
\end{split}
\label{eq:kingen6}
\end{equation}
where $|\mathbf{\bar{p}}_{i}|$ is the $i$-th particle momentum modulus, determined by energy-momentum 
conservation, in the $i$-th 2-body decay rest frame, namely 

\begin{equation}
|\mathbf{\bar{p}}_{i}| = \displaystyle\frac{{\left[\left(q_{i-1}^{2}-\left(m_{i}+q_{i}\right)^{2}\right)
\left(q_{i-1}^{2}-\left(m_{i}-q_{i}\right)^{2}\right)\right]}^{1/2}}{2q_{i-1}},
\label{eq:kingen7}
\end{equation}
for $i=1,...,N-2$, and 

\begin{equation}
\begin{split}
&|\mathbf{\bar{p}}_{i}| =\\
&\displaystyle\frac{{\left[\left(q_{N-2}^{2}-\left(m_{N-1}+m_{N}\right)^{2}\right)
\left(q_{N-2}^{2}-\left(m_{N-1}-m_{N}\right)^{2}\right)\right]}^{1/2}}{2q_{N-2}}
\end{split}
\label{eq:kingen8}
\end{equation}
for $i=N-1$ and $i=N$. Moreover, the energy components $\bar{p}_{i}(0)$ and $\bar{q}_{i}(0)$ are given
by

\begin{equation}
\begin{split}
&\bar{p}_{i}(0) = \sqrt{m_{i}^{2} + |\mathbf{\bar{p}}_{i}|^{2}}\\
&\bar{q}_{i}(0) = \sqrt{q_{i}^{2} + |\mathbf{\bar{p}}_{i}|^{2}}\\
\end{split}
\end{equation}
and $E_{i}$ is the $i$-th particle energy in the cluster rest frame corresponding to the momentum $\mathbf{\bar{p}}_{i}$.

It follows from Eq. \ref{eq:kingen6} that the needed $N$-body phase space configuration can be obtained by sampling
$N$-1 solid angles, corresponding to the particle emission direction in the 2-body decays, and $N$-2 virtualities. 
In the present work, the solid angles $\Omega_{\mathbf{\bar{p}}_{i}}$ and the $q_{i}^{2}$ vir\-tualities are randomly 
generated within $4\pi$ and within the limits defined in Eq. \ref{eq:kingenlim} respectively. The particle momenta 
generated with this procedure, initially defined in the 2-body decay rest frames, are subsequently boosted to the 
laboratory frame. Finally, the phase space sampling weight $w_{PS}$ is given by

\begin{equation}
\begin{split}
w_{PS} &= \frac{(4\pi)^{N-1}}{2^{N-2}}\displaystyle\prod_{i=1}^{N-2}\frac{|\mathbf{\bar{p}}_{i}|}{\bar{p}_{i}(0)+\bar{q}_{i}(0)}
\left(q_{i}^{2\ max}-q_{i}^{2\ min}\right)\times\\
&\frac{|\mathbf{\bar{p}}_{N-1}|}{\bar{p}_{N-1}(0)+\bar{p}_{N}(0)}\displaystyle\prod_{i=1}^{N}E_{i},
\end{split}
\label{eq:kingen9}
\end{equation}

Defining the hadronization channel sampling weight $w_{HC}$ as

\begin{equation}
w_{HC} = \displaystyle\prod_{j=1}^{K}\frac{1}{N_{j}!}\left[\frac{\left(2J_{j}+1\right)V}{\left(2\pi\right)^{3}}\right]^{N_{j}}
\frac{1}{\Pi_{\{N_{j}\}}}
\label{eq:hcweight1}
\end{equation}
for light clusters and as

\begin{equation}
w_{HC} = \displaystyle\prod_{j=1}^{K}\frac{1}{N_{j}!}\left[\frac{\left(2J_{j}+1\right)V}{\left(2\pi\right)^{3}}\right]^{N_{j}}
\frac{1}{P_{h}\Pi_{\{N_{j}\}}}
\label{eq:hcweight2}
\end{equation}
for heavy flavored clusters, where the hadronization channel sampling functions $\Pi_{\{N_j\}}$ and $P_{h}$ 
are defined in Eqs. \ref{eq:samplf1} and \ref{eq:heavysampl} respectively, the total weight $w$ corresponding 
to a single cluster hadronization event reads 

\begin{equation}
\begin{split}
w = \frac{w_{HC}w_{PS}}{\Omega}.
\end{split}
\label{eq:totalsingle}
\end{equation}

\section{Comparison with the Cluster Model}
\label{sec:StdCluMod}

Although the SHM is based on the general idea of cluster hadronization, it has peculiar
differences with respect to other hadronic event generators implementing Cluster Model
hadronization such as \texttt{Herwig++} \cite{HerwigNew} and \texttt{Sherpa} \cite{Sherpa}, 
besides \texttt{Herwig6510}. Let us now discuss the diffe\-rences between these approaches 
more in detail starting from the \texttt{Herwig} case. 

The most striking difference is concerned with the generation of the channel multiplicity: 
whilst the Cluster Model forces clusters to decay into hadron pairs, in SHM the final 
multiplicity is an outcome of the cluster hadronization process. Particularly, the relative 
production rate of $N$-body channels is determined by the cluster's finite volume. 
This happens because the SHM uses the proper measure of the multihadronic phase space 
\cite{becareview,BecMeaning} with a cluster's proper volume $V$ acting as a coupling constant 
for the cluster decay according to Eq.~\ref{eq:microweigth}. Conversely, the Cluster Model 
uses invariant momentum space measure $\mathrm{d}^3 p/2 E$ without extra factors and therefore the 
final multiplicity is to be fixed otherwise (see extensive discussion in \cite{BecMeaning}). 

Indeed, in the \texttt{Herwig} event generator, a hadronizing cluster can be transformed 
in three different ways depending on its mass:
\begin{itemize} 
\item splitting into a lighter cluster pair (see discussion in Sec. \ref{subsec:HWClusterSplit});
\item decay into a hadron pair;
\item transformation into the lightest hadron with the same flavor composition of the 
cluster.
\end{itemize}
Here we point out a further difference, even for the two-body decays: given a cluster with flavor 
composition ${\rm f}_{1}{\rm \bar{f}}_{2}$, in the \texttt{Herwig} Cluster Model the decay hadron pair 
will have the flavor composition ${\rm f}_{1}{\rm \bar{f}}$ and ${\rm f}{\rm \bar{f}}_{2}$ 
respectively, with the hadrons sharing the components of the ${\rm f}{\rm \bar{f}}$ parton pair extracted
from the vacuum to perform the cluster decay in this framework. On the other hand, in the SHM 
only abelian charges conservation is required, implying the inclusion of a larger set of 
channels. 

With respect to the general aforementioned \texttt{Herwig} hadronization scheme, 
\texttt{Sherpa} implementation of the Cluster Model features additional differences with respect 
to SHM. Indeed, relevant changes have been introduced in implementing the Cluster Model in the event 
generator \cite{SherpaClu}, both for cluster production and decay: inclusion of soft color reconnection 
during cluster formation and decay; inclusion of the parton/hadron spin degree of freedom in the transition 
probability calculation; a refinement of the single hadron cluster transformation. Particularly, in the 
\texttt{Sherpa} case all clusters with mass falling into the so-called hadronic regime are transformed 
into single hadrons with the same flavor compositions of the hadronizing clusters. Each specific hadron 
is then randomly chosen among those with a mass smaller than the one of the cluster itself, instead of 
always choosing the lightest one as in the \texttt{Herwig} case.

\section{Multi-cluster hadronization and event weight}
\label{sec:MultiClusterAlgo}

Within cluster hadronization models, high energy \ee \\collisions (say $\sqrt s \ge 4\ 
\textnormal{GeV}$) always involve multiple cluster production. For instance, the Cluster Model 
\cite{ClusterH} implemented in \texttt{Herwig6510} event generator \cite{HerwigOld}, provides
the formation of color connected parton-anti\-parton pairs at the end of the perturbative 
QCD shower, gi\-ving rise to massive colorless clusters. As discussed in Sec. \ref{sec:StdCluMod}, those clusters 
are thereafter decayed into pairs of massive hadrons, the probability of the specific decay channel 
being determined by the two-body phase space availability and by the spin multiplicities of the involved 
hadrons. Indeed, in the present work, the multi-cluster system formed as a result of the hard 
scattering process, followed by the perturbative showering, is taken from the \texttt{Herwig6510} event 
generator and used as input for the SHM. Speci\-fically, the general simulation setup is as follows: the 
\texttt{Herwig6510} event generator is used to simulate the full pre-hadronization process, 
from initial state radiation emission to cluster production. The \texttt{Herwig6510} preliminary cluster 
splitting procedure described in \cite{HerwigOld}, which follows the cluster formation step, is also included in 
the event simulation (the details of the cluster splitting procedure and its impact on 
the SHM predictions are discussed in Sec. \ref{subsec:HWClusterSplit}). At that stage, the standard 
hadronization algorithm is replaced by the SHM one, which takes care of hadronizing all produced clusters;
finally, the formed hadrons are returned to the external generator (i.e. \texttt{Herwig6510}) 
which takes care of decaying of the unstable particles.  

The above hadronization procedure is applied to each cluster independently, thus, for each 
simulated collision, the hadronization event weight $W$ reads:

\begin{equation}
W = \prod_{i=1}^{N_{\textnormal{C}}}w_{i}=\frac{\displaystyle\prod_{i=1}^{N_{\textnormal{C}}}w^{i}_{HC}w^{i}_{PS}}
{\displaystyle\prod_{i=1}^{N_{\textnormal{C}}}\Omega_{i}}, 
\label{eq:hadrototweigth}
\end{equation} 
where $N_{\textnormal{C}}$ is the number of clusters in the event and $w_{i}$ the $i$-th cluster 
hadronization weight defined in Eq. \ref{eq:totalsingle}. Moreover, $w^{i}_{HC}$ and $w^{i}_{PS}$ 
are the hadronization channel and phase space sampling weights of the $i$-th cluster 
(defined in Eqs. \ref{eq:kingen9}, \ref{eq:hcweight1} and \ref{eq:hcweight2}) and $\Omega_{i}$ 
its partition function (defined in Eq. \ref{eq:micropart}). 

This method in fact involves a complication, related to the calculation of the hadronization
event weight of Eq. \ref{eq:hadrototweigth}. Specifically, it requires the prior knowledge of 
each cluster's microcanonical partition function. If we had only single cluster event with fixed 
mass and charges, its microcanonical partition function would be an irrelevant constant factor 
cancelling out when calculating mean values. 
On the other hand, for a multi-cluster environment, the value of the product of microcanonical partition
functions in Eq. \ref{eq:hadrototweigth} is not constant, in fact it is a function of the masses and the charges 
of the particular set of clusters. Therefore, we need to know the specific numerical value
of the microcanonical partition functions of the clusters produced in a single event in order
to correctly normalize its weight.

Unfortunately, computing the microcanonical partition function of a cluster of given mass and
set of charges is a CPU time demanding task that cannot be afforded at the event generation time.
To solve this problem, we decided to pre-calculate microcanonical partition functions for
a discrete set of masses (and charges) and store their values in a look-up table to be used at 
event generation time. The number of pre-calculated functions has been determined on the basis
of the observation that, in about $10^{6}$ \ee collision events at the Z peak, around 150 different 
cluster charge configurations (including baryon number, strangeness, electric charge, charm 
and beauty) occurred. For each of these charge configurations, a set of microcanonical partition 
functions has been calculated, each set including different cluster mass and different free 
parameters ($\gamma_{\textnormal{S}}$ and $\rho$) values. Specifically, for each charge configuration, 
a grid in the mass-$\rho$-$\gamma_S$ space has been defined as follows:
\begin{itemize}
\item $\rho \in [0.20;0.50]$ GeV/$\textnormal{fm}^{3}$ with $\Delta\rho = 0.05$ GeV/$\textnormal{fm}^{3}$; 
\vspace{0.05 mm}
\item $\gamma_{\textnormal{S}} \in [0.50;1.00]$ with $\Delta\gamma_{\textnormal{S}} = $ 0.05;
\vspace{4. mm}
\item Cluster mass $M \in [2m_{\pi};10\ \textnormal{GeV}]$ with:
	\begin{itemize}
 	\item $\Delta M = $ 0.1 GeV for $M\leq3$ GeV;
	\item $\Delta M = $ 0.5 GeV otherwise;
	\end{itemize}
\end{itemize}
and for each grid point an $\Omega$ calculated. Then, during event generation, the partition 
function values of the single clusters, for their specific charge configuration/mass and for 
the chosen SHM free parameter values, are determined by means of a linear interpolation between
the nearest points in the grid. 

As it can be seen, a non uniform grid structure along the cluster mass direction has been adopted: 
this choice was motivated by the non-smooth dependence of the microcanonical partition function
on the mass at low mass values. Indeed, $3$ GeV as mass value separating the two 
differently discretized regions proved to be a good trade-off between the need of good accuracy
and the computational cost in getting all grid points calculated.

A final remark concerning the calculation of the partition function set is in order: 
for a given mass, charge configuration and $\gamma_{\textnormal{S}}$ and $\rho$
parameters values, the corresponding microcanonical partition function value is determined 
by the set of hadrons used in the hadronization process and by their physical properties. 
It follows that the set of partition functions needs to be rebuilt only when changes are introduced in 
the hadron set, such as modifications to the hadron physical properties or to the hadron list.

The details of the numerical methods used for the pre-calculation of microcanonical partition 
function grid can be found in \cite{phdthesis}. It is worth mentioning here that we have used
the phase space integration optimized algorithm described in \cite{BecFerr3,WernAich}.

\section{Preliminary tests}
\label{sec:SHMPreliminaryTest}

In this section the results obtained with the SHM code, interfaced to \texttt{Herwig6510}, 
for \ee collisions at $91.2$ GeV center of mass energy, will be presented and discussed. 
While a rigorous tuning of the SHM free para\-meters is needed to assess the 
goodness of the model, the results presented here, obtained with an approximate parameter
adjustment, already show a good accuracy level. Moreover, in view of a global tuning of 
both SHM and \texttt{Herwig6510} free parameters, a preliminary analysis of the
dependency of SHM predictions on some \texttt{Herwig6510} parameters involved in the
cluster formation step is presented.

\subsection{Simulation setup}

As already discussed, the \texttt{Herwig6510} event generator has been used for the present work as external 
code for the collision simulation, replacing its standard hadronization routines with the SHM 
Monte-Carlo module. As input clusters for the microcanonical hadronization, the clusters normally 
produced by \texttt{Herwig6510} have been used, except for two modifications described below. 
The set of hadrons produced during each microcanonical hadronization process is then processed by 
\texttt{Herwig6510}, which performs the unstable particle decays.

The modifications introduced in the standard cluster sets produced by \texttt{Herwig6510} during the event
simulation, and above mentioned, are the following:

\begin{itemize}
\item Cluster merging: the predictions of the SHM show a dependency on the input cluster 
mass spectrum, which in its turn depends on the \texttt{Herwig6510} setup. With the final aim of obtaining a 
better agreement between the SHM predictions and the experimental data, it is useful to have the possibility 
to modify the properties of the cluster mass spectrum. In the present work, the above modification is introduced 
by means of a cluster merging procedure, which, working after the \texttt{Herwig6510} cluster production
step, allows to control their mass spectrum without any changes in the external generator standard setup.
Starting from a set of  \texttt{Herwig6510} clusters, the result of the merging procedure is determined by 
the value of a free parameter ($M_{\textnormal{C}}$) representing the minimum allowed cluster mass: the set of 
incoming clusters is analyzed in order to check the presence of objects with mass under the given threshold. 
If one or more light clusters are found, an iterative fusion process is activated, which merges the cluster pairs 
containing at least one light object into heavier clusters, repeating this procedure until no light clusters remain. 
When more than one combination for a light cluster is possible the merged pair is the one with the smallest 
invariant mass
 
\begin{equation}
m(i,j) = \sqrt{\left(p_{i}+p_{j}\right)^{2}},
\label{eq:invmass}
\end{equation}
where $p_{i}(p_{j})$ is the 4-momentum of the $i$-th($j$-th)  cluster. It must be noted that a maximum mass 
value of 10 GeV is allowed for the new clusters produced during the merging procedure, a condition required 
to control the broadening of the cluster mass spectrum towards large mass values and the consequent quick 
increase in the computational cost of the corresponding partition function grid calculation. Because of the above condition, 
when a cluster pair is chosen for the merging procedure, the corresponding new cluster is created only if the pair 
invariant mass (Eq. \ref{eq:invmass}) is smaller than 10 GeV: in a limited number of cases this 
condition can prevent the cluster merging procedure from being fully completed, with the result that clusters with 
mass below the chosen $M_{\textnormal{C}}$ value can still be included in the hadronization procedure.

\item Baryonic clusters: the second change to the standard \texttt{Herwig6510} cluster properties is related to 
the production of baryonic clusters, disabled in a standard run of this generator, and is obtained by means 
of a proper setting of its \texttt{QDIQK} parameter value. This parameter represents the maximum scale at which gluons 
can be (non-perturbatively) split into diquarks during the QCD shower process and is set to zero in the default \texttt{Herwig6510} 
configuration, thus switching off the possibility of diquarks production. For the present work the 
value $2m_{\rm c} = 2\times1.55$ GeV has been chosen for the \texttt{QDIQK} parameter, 
with the result of activating the possibility of gluon splitting into light flavored diquark-antidiquark pairs. During 
the subsequent \texttt{Herwig6510} cluster building step, each produced diquark (antidiquark) is coupled to the corresponding 
color connected quark (antiquark) to form a colorless baryonic cluster. The need for baryonic clusters is strictly 
related to the correct baryon production during the microcanonical hadronization process: because of the 
baryonic charge conservation, baryons can be obtained from the standard \texttt{Herwig6510} non-baryonic 
clusters only as baryon-antibaryon pairs. However, these configurations are strongly suppressed because
of phase space availability reasons. The introduction of baryonic clusters, on the other side, allows to obtain 
the production of baryonic final states in less restrictive phase space conditions.

\end{itemize}

\begin{figure}[!h]
\includegraphics[width=0.450\textwidth]{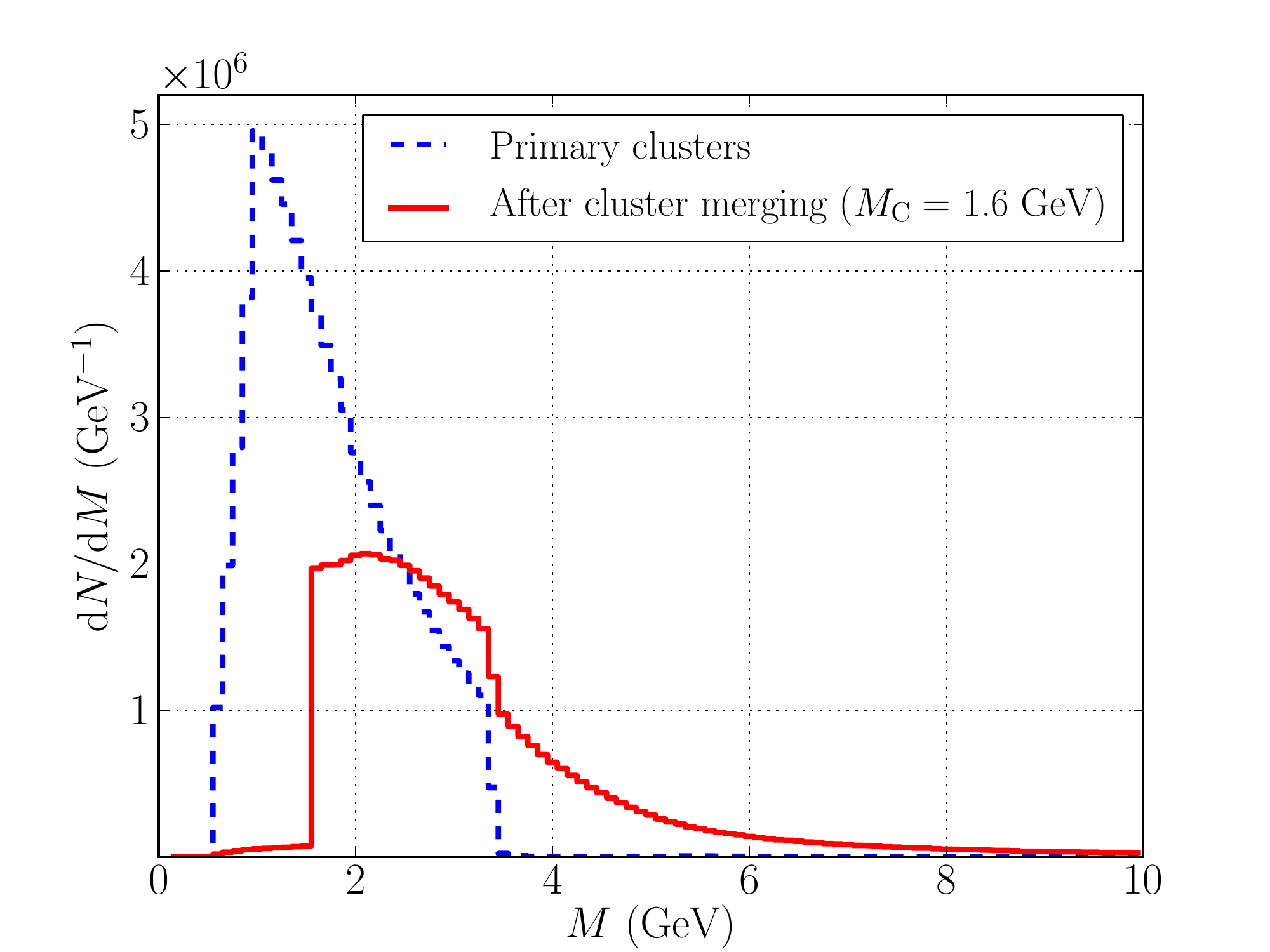}
\caption[Cluster invariant mass distribution for $10^{6}$ \mbox{${\rm e}^+{\rm e}^- \rightarrow {\rm d} \bar{{\rm d}}$} 
events at 91.2 GeV: comparison among primary clusters and clusters obtained with the merging procedure described in 
the text.]{Cluster invariant mass distribution for $10^{6}$ \mbox{${\rm e}^+{\rm e}^- \rightarrow {\rm d} \bar{{\rm d}}$} 
events at 91.2 GeV: comparison among primary clusters and clusters obtained with the merging procedure described in
the text.}
\label{fig:m_clusterNotNormDD}
\end{figure}

In Fig. \ref{fig:m_clusterNotNormDD} the effects of the discussed cluster merging procedure on the cluster invariant mass 
spectrum are shown, in particular for clusters produced in $10^{6}$ \mbox{${\rm e}^+{\rm e}^- \rightarrow {\rm d} \bar{{\rm d}}$} 
collisions at 91.2 GeV. Primary clusters are the standard \texttt{Herwig6510} clusters, except for the activation of baryonic
cluster production, and their mass spectrum is compared with the one of clusters obtained with the merging procedure
for $M_{\textnormal{C}}=$ 1.6 GeV: the presence of the cluster minimum mass selection is evident, as well as of a broadening 
of the mass distribution. At the same time, it can be seen that a residual fraction of clusters, about 1\%, have a mass below 
the $M_{\textnormal{C}}$ value: the existence of these objects after the merging procedure is due to the 
maximum allowed cluster mass value of the merging procedure which, as already discussed, in a limited number 
of cases prevents the cluster merging from being completed. The mass distribution of the secondary clusters, namely the
ones provided in output by the merging procedure, shows also an inflection point around 3.5 GeV. As it can be seen in 
Fig. \ref{fig:m_clusterNotNormDD}, this mass value is the maximum allowed for primary clusters, a condition responsible for the
observed inflection point: in fact, only new clusters produced by the merging procedure contributes to the mass distribution for 
values larger than the above limit, while for smaller mass values contributions from the primary cluster 
mass distribution are also present. These differences in the secondary cluster mass distribution composition, in the two mass
ranges separated by the primary cluster maximum mass value, is the origin of the beha\-vior change in the 
mass distribution. 

Finally, we point out that in the hadronization step, among the light flavored hadron states, only those with 
mass less or equal to 1.8 GeV were included. This choice was motivated by the opportunity of comparing 
multiplicities with previous results obtained with the SHM where this cutoff was used. It should be noted, 
anyhow, that the production of light flavored states with mass larger than 1.8 GeV is negligible for essentially 
all observables.

\subsection{Study of \texttt{Herwig} cluster splitting}
\label{subsec:HWClusterSplit}

As already stated, the \texttt{Herwig6510} cluster splitting procedure described in \cite{HerwigOld} is included in the present event
simulation setup. The splitting procedure, which in  \texttt{Her\-wig6510} precedes the cluster decay step, operates 
as follows: a cluster of mass $M_{{\rm f}_{1}\bar{\rm f}_{2}}$, composed of the partons ${\rm f}_{1}$ and 
$\bar{\rm f}_{2}$ of mass $m_{{\rm f}_{1}}$ and $m_{\bar{\rm f}_{2}}$ respectively, is split if the following 
condition holds:
\begin{equation}
\left(M_{{\rm f}_{1}\bar{\rm f}_{2}}\right)^{{\rm \texttt{CLPOW}}} \geq {\rm \texttt{CLMAX}}^{{\rm \texttt{CLPOW}}} + \left(m_{{\rm f}_{1}}+
m_{\bar{\rm f}_{2}}\right)^{{\rm \texttt{CLPOW}}},
\label{eq:splitcondition}
\end{equation}
where \texttt{CLPOW} and \texttt{CLMAX} are phenomenological parameters.
For each cluster satisfying the above condition a flavor pair ${\rm f}\bar{\rm f}$ is picked from the vacuum
and its components used for the construction of the two daughter clusters, whose flavor compositions
will be ${\rm f}_{1}\bar{\rm f}$ and ${\rm f}\bar{\rm f}_{2}$ respectively. The masses $M_{{\rm f}_{1}\bar{\rm f}}$ and 
$M_{{\rm f}\bar{\rm f}_{2}}$ of the newly created clusters are randomly generated according to the equations
\begin{equation}
\begin{split}
& M_{{\rm f}_{1}\bar{\rm f}} = m_{{\rm f}_{1}} + (M_{{\rm f}_{1}\bar{\rm f}_{2}} - m_{{\rm f}_{1}} - m_{{\rm f}})r^{1/\mathrm{\texttt{PSPLT}}} \\
& M_{{\rm f}\bar{\rm f}_{2}} = m_{\bar{\rm f}_{2}} + (M_{{\rm f}_{1}\bar{\rm f}_{2}} - m_{\bar{\rm f}_{2}} - m_{{\rm f}})r^{1/\mathrm{\texttt{PSPLT}}}
\end{split}
\end{equation}
where $m_{{\rm f}}$ is the mass of the generated parton pair components, $r$ is an uniformly 
distributed random number and where $\texttt{PSPLT}$ is a third free parameter of the splitting 
procedure (in fact, \texttt{PSPLT} is a two component parameter: one component refers to light 
flavored and charmed clusters and the other to the bottomed ones). 

\begin{figure}
\includegraphics[width=0.450\textwidth]{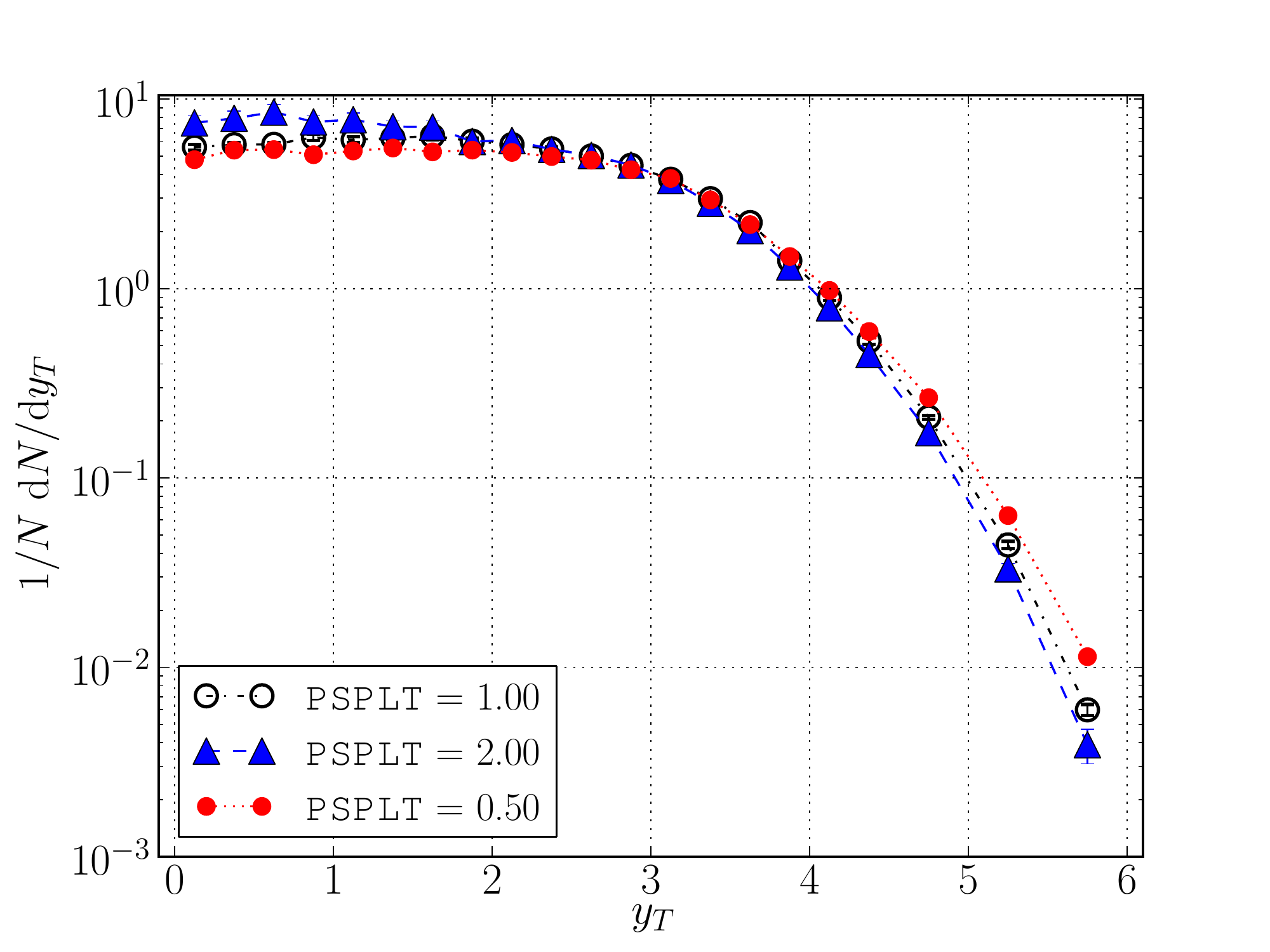}
\caption[Rapidity with respect to thrust axis normalized distribution: \texttt{PSPLT} parameter variation effects on SHM predictions for \mbox{${\rm e}^+{\rm e}^- \rightarrow {\rm d} \bar{{\rm d}}$} collisions at $91.2$ GeV center of mass energy.]{Rapidity with respect to thrust axis normalized distribution: \texttt{PSPLT} parameter variation effects on SHM predictions for \mbox{${\rm e}^+{\rm e}^- \rightarrow {\rm d} \bar{{\rm d}}$} collisions at $91.2$ GeV center of mass energy.}
\label{fig:y_TPsplt}
\end{figure}

\begin{figure}
\includegraphics[width=0.450\textwidth]{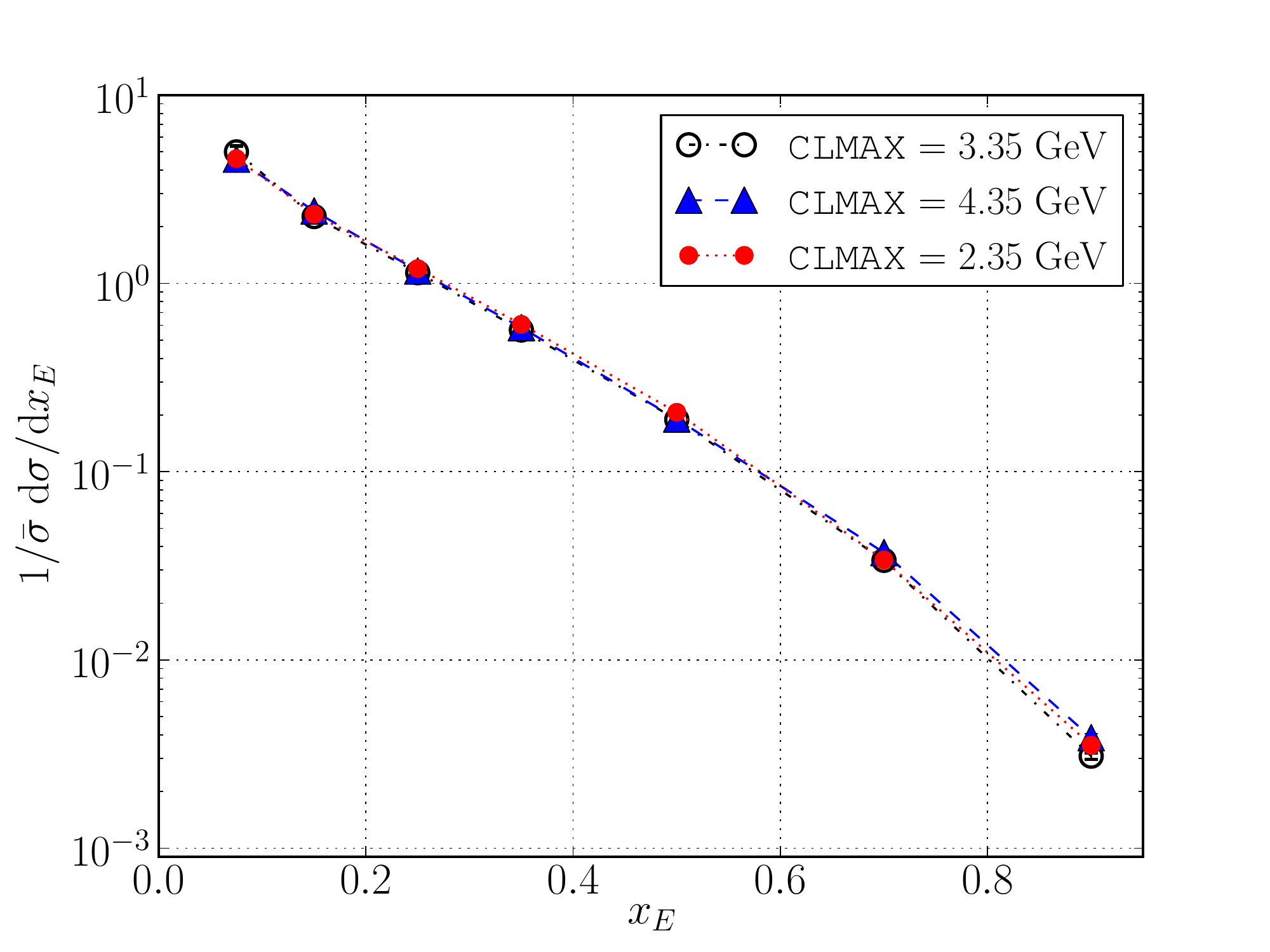}
\caption[$\rho^{0}$ scaled energy distribution: \texttt{CLMAX} parameter variation effects on SHM predictions for \mbox{${\rm e}^+{\rm e}^- \rightarrow {\rm d} \bar{{\rm d}}$} collisions at $91.2$ GeV center of mass energy.]{$\rho^{0}$ scaled energy distribution: \texttt{CLMAX} parameter variation effects on SHM predictions for \mbox{${\rm e}^+{\rm e}^- \rightarrow {\rm d} \bar{{\rm d}}$} collisions at $91.2$ GeV center of mass energy.}
\label{fig:x_e_rhozeroClmax}
\end{figure}

\begin{figure}
\includegraphics[width=0.450\textwidth]{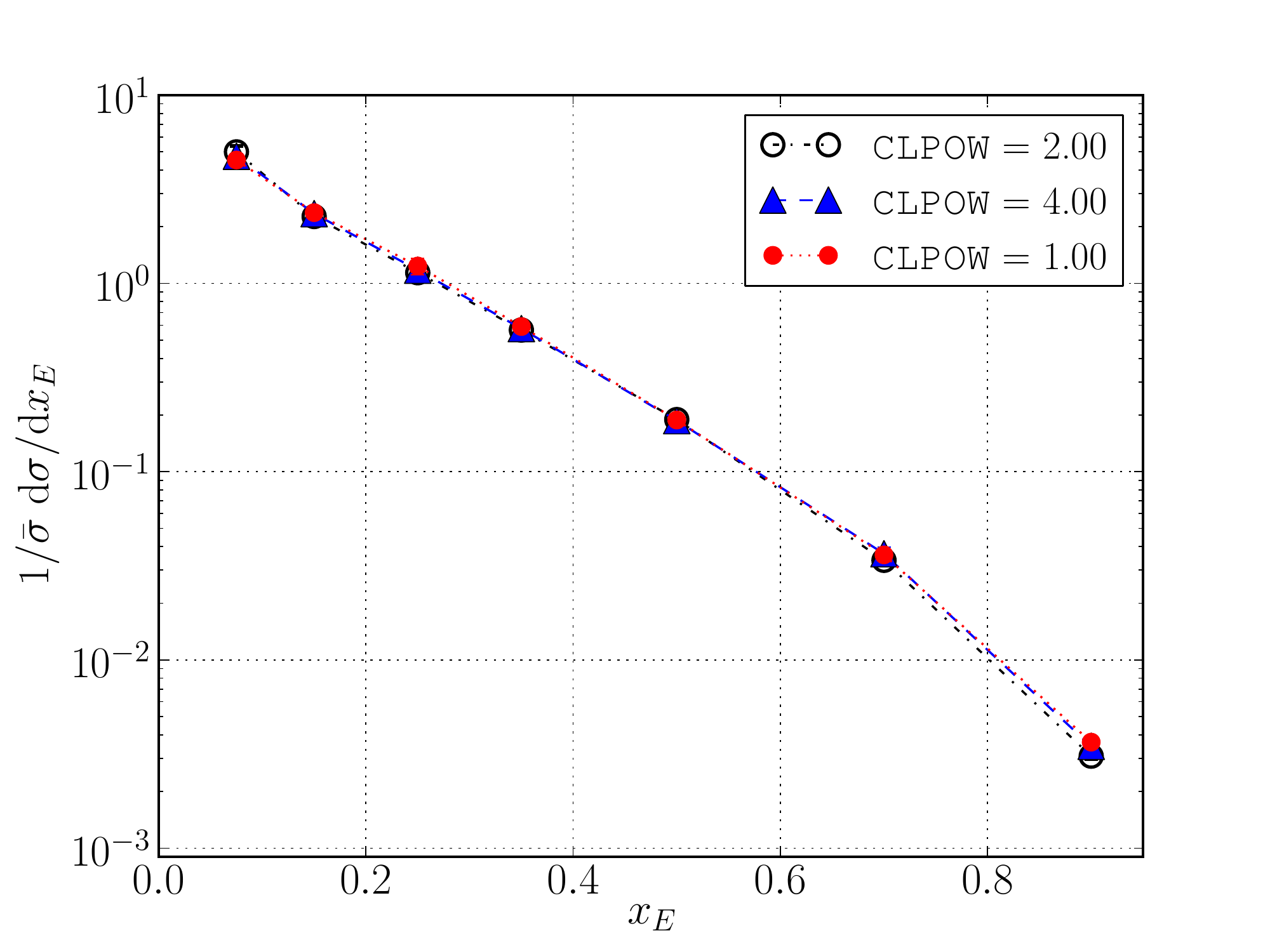}
\caption[$\rho^{0}$ scaled energy distribution: \texttt{CLPOW} parameter variation effects on SHM predictions for \mbox{${\rm e}^+{\rm e}^- \rightarrow {\rm d} \bar{{\rm d}}$} collisions at $91.2$ GeV center of mass energy.]{$\rho^{0}$ scaled energy distribution: \texttt{CLPOW} parameter variation effects on SHM predictions for \mbox{${\rm e}^+{\rm e}^- \rightarrow {\rm d} \bar{{\rm d}}$} collisions at $91.2$ GeV center of mass energy.}
\label{fig:x_e_rhozeroClpow}
\end{figure}

\begin{figure}
\includegraphics[width=0.450\textwidth]{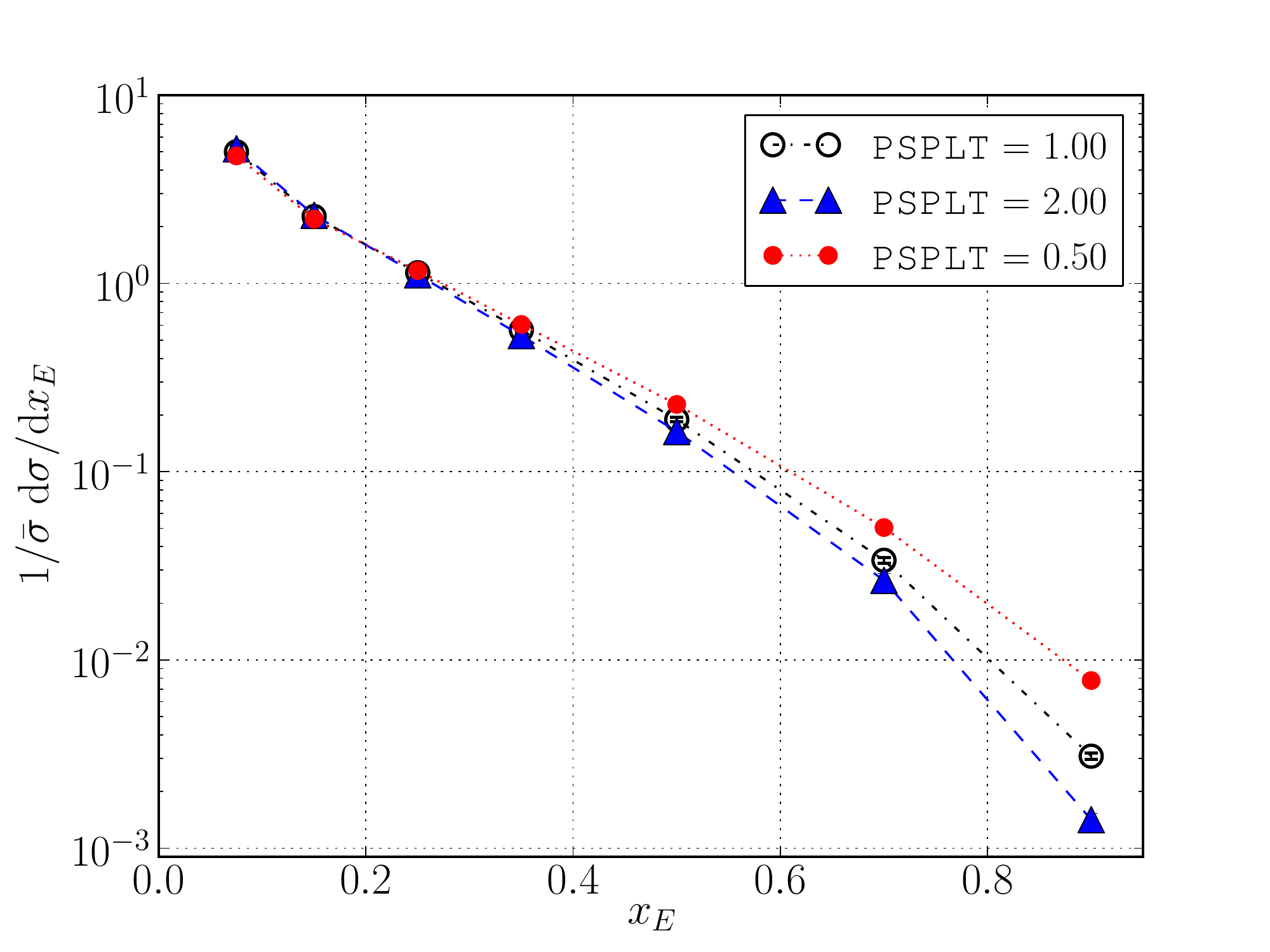}
\caption[$\rho^{0}$ scaled energy distribution: \texttt{PSPLT} variation effects on SHM predictions for \mbox{${\rm e}^+{\rm e}^- \rightarrow {\rm d} \bar{{\rm d}}$} collisions at $91.2$ GeV center of mass energy.]{$\rho^{0}$ scaled energy distribution: \texttt{PSPLT} parameter variation effects on SHM predictions for \mbox{${\rm e}^+{\rm e}^- \rightarrow {\rm d} \bar{{\rm d}}$} collisions at $91.2$ GeV center of mass energy.}
\label{fig:x_e_rhozeroPsplt}
\end{figure}

While a detailed analysis of the interplay between SHM free parameters and those belonging to 
\texttt{Herwig6510} initial collision steps is needed in view of a global tuning, especially concerning
the cluster formation and splitting free parameters such as \texttt{CLPOW}, \texttt{CLMAX} and \texttt{PSPLT}, 
a preliminary evaluation of their impact on the SHM predictions is here reported. The discussed numerical 
results have been obtained by means of the following single variations in the above cluster splitting parameter 
default values (the default condition $\texttt{PSPLT(2)} = \texttt{PSPLT(1)}$ has been used in the 
present analysis):
\begin{itemize}
\item $\texttt{CLMAX}$ = 2.35 - 3.35 (default) - 4.35 GeV;
\item $\texttt{CLPOW}$ = 1.00 - 2.00 (default) - 4.00;
\item $\texttt{PSPLT}$ = 0.50 - 1.00 (default) - 2.00.
\end{itemize}
The analysis has been focused on \mbox{${\rm e}^+{\rm e}^- \rightarrow {\rm d} \bar{{\rm d}}$} collisions at 91.2 GeV 
and the following values have been used for the SHM hadronization module free parameters:
\begin{itemize}
\item $M_{\textnormal{C}} = 1.6$ GeV;
\item $\rho = 0.45$  GeV/$\textnormal{fm}^{3}$;
\item $\gamma_{\textnormal{S}} = 0.65$.
\end{itemize} 
The obtained results show that only small effects are introduced by the considered variations 
in the $\texttt{CLMAX}$ and $\texttt{CLPOW}$ parameter values, as can be seen e.g. for 
the rapidity distribution reported in Fig. \ref{fig:y_TClmax} and \ref{fig:y_TClpow}
(for the definition of the observables, see Appendix). On the other side a considerable dependency on the $\texttt{PSPLT}$ 
parameter value is observed in the SHM predictions, as clearly visible in Fig. \ref{fig:y_TPsplt} again for event rapidity. 
The same situation is present for single particle observables, as reported for example in Figs. \ref{fig:x_e_rhozeroClmax} - 
\ref{fig:x_e_rhozeroPsplt} for $\rho^{0}$ meson scaled energy distribution, and for particle multiplicities 
(Tabs. \ref{tab:multClmax} - \ref{tab:multPsplt}). Concerning multiplicities, it is worth noting the large variations in the total 
number of charged particles and in light meson multiplicities due to the modification of the $\texttt{PSPLT}$ parameter, 
while $\texttt{CLMAX}$ and $\texttt{CLPOW}$ variation effects are essentially limited to the baryonic sector.

\begin{figure}
\includegraphics[width=0.450\textwidth]{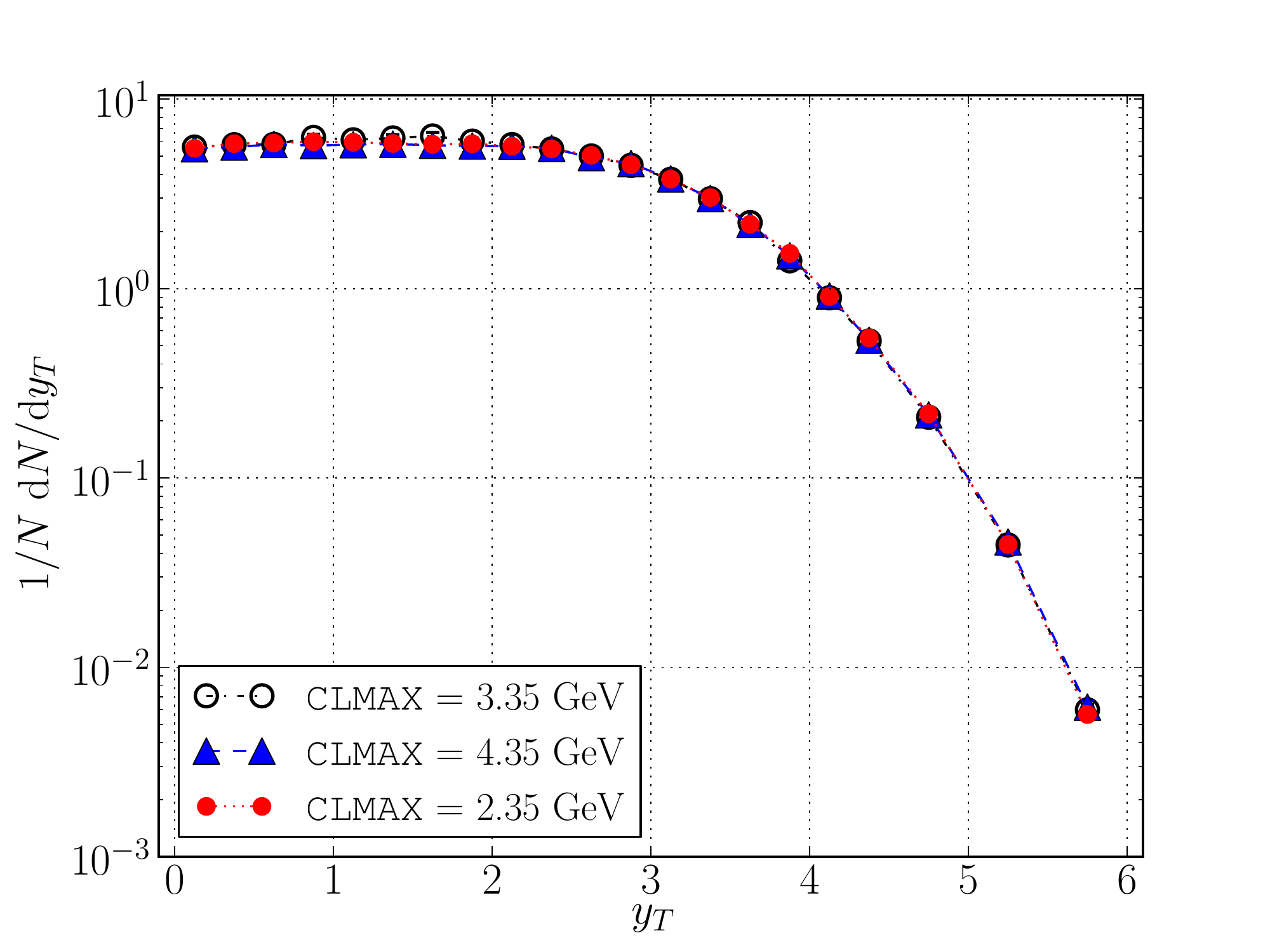}
\caption[Rapidity with respect to thrust axis normalized distribution: \texttt{CLMAX} parameter variation effects on SHM predictions for \mbox{${\rm e}^+{\rm e}^- \rightarrow {\rm d} \bar{{\rm d}}$} collisions at $91.2$ GeV center of mass energy.]{Rapidity with respect to thrust axis normalized distribution: \texttt{CLMAX} parameter variation effects on SHM predictions for \mbox{${\rm e}^+{\rm e}^- \rightarrow {\rm d} \bar{{\rm d}}$} collisions at $91.2$ GeV center of mass energy.}
\label{fig:y_TClmax}
\end{figure}

\begin{figure}
\includegraphics[width=0.450\textwidth]{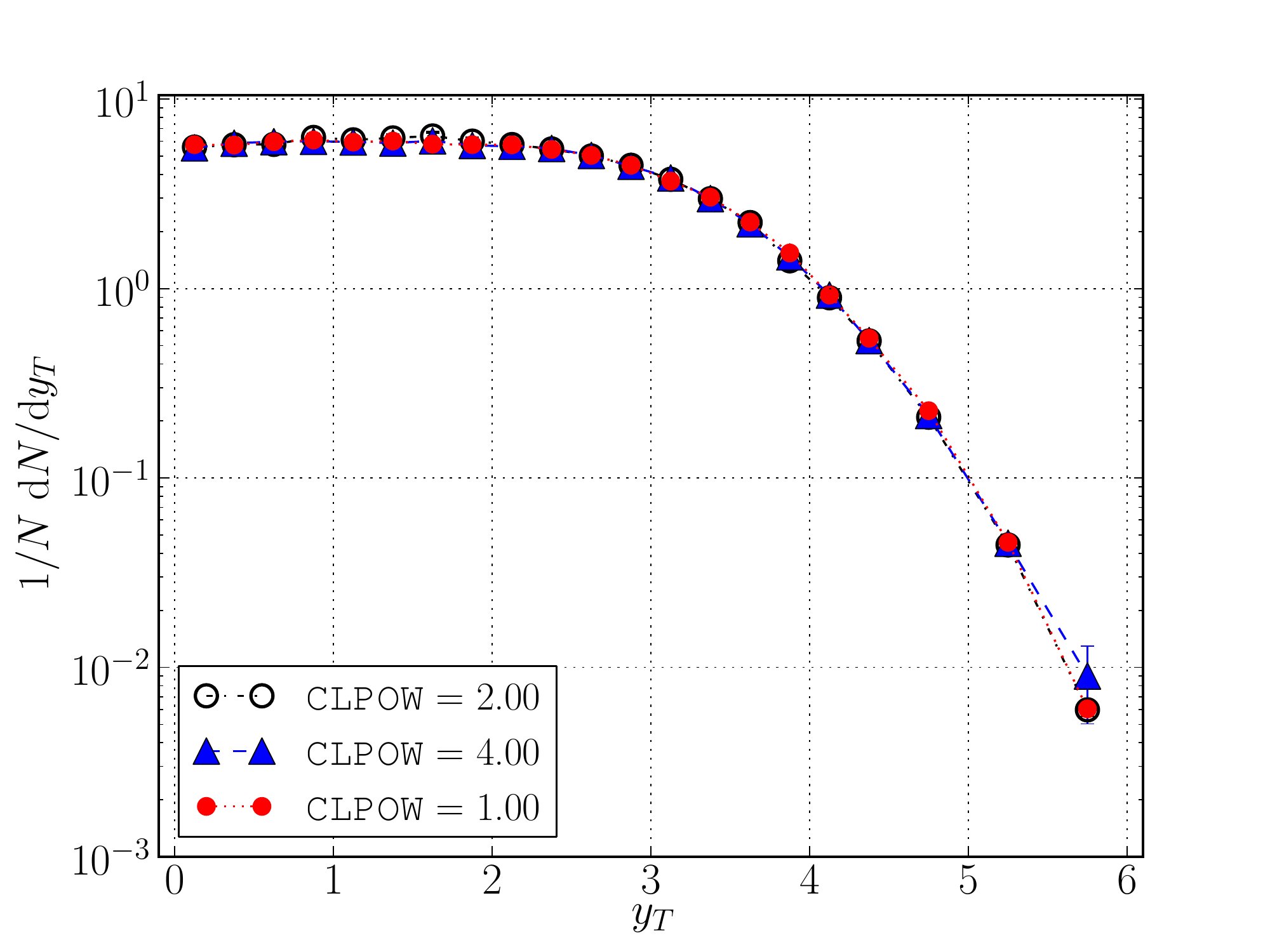}
\caption[Rapidity with respect to thrust axis normalized distribution: \texttt{CLPOW} parameter variation effects on SHM predictions for \mbox{${\rm e}^+{\rm e}^- \rightarrow {\rm d} \bar{{\rm d}}$} collisions at $91.2$ GeV center of mass energy.]{Rapidity with respect to thrust axis normalized distribution: \texttt{CLPOW} parameter variation effects on SHM predictions for \mbox{${\rm e}^+{\rm e}^- \rightarrow {\rm d} \bar{{\rm d}}$} collisions at $91.2$ GeV center of mass energy.}
\label{fig:y_TClpow}
\end{figure}

\newpage

\begin{table*}
\begin{center}
\caption[Charged particles and hadron multiplicity mean values: \texttt{CLMAX} parameter variation effects on SHM predictions for \mbox{${\rm e}^+{\rm e}^- \rightarrow {\rm d} \bar{{\rm d}}$} collisions at $91.2$ GeV center of mass energy.]{Charged particles and hadron multiplicity mean values: \texttt{CLMAX} parameter variation effects on SHM predictions for \mbox{${\rm e}^+{\rm e}^- \rightarrow {\rm d} \bar{{\rm d}}$} collisions at $91.2$ GeV center of mass energy.}
\label{tab:multClmax}
\begin{tabular}{c c c c}
\hline\noalign{\smallskip}
 & Default & \texttt{CLMAX} = 2.35 GeV & \texttt{CLMAX} = 4.35 GeV \\ 
\noalign{\smallskip}\hline\noalign{\smallskip}
Charged		&	$20.29 \pm 0.32$		&$19.93 \pm 0.10$		&$19.64 \pm 0.17$\\		
$\pi^{0}$		&	$10.09 \pm 0.08$		&$10.10 \pm 0.06$		&$9.95 \pm 0.09$\\		
$\pi^{+}$		&	$8.71 \pm 0.16$		&$8.55   \pm 0.05$   		&$8.46 \pm 0.08$\\		
$\eta$		&	$1.24 \pm 0.03$		&$1.25   \pm  0.02$		&$1.21 \pm 0.02$\\		
$\Delta^{++}$	&	$0.081 \pm 0.002$		&$0.0724 \pm 0.0007$	&$0.080 \pm 0.002$\\		
$\Sigma^{+}$	&	$0.0277 \pm 0.0006$	&$0.0245 \pm 0.0003$	&$0.0229 \pm 0.0004$ \\ 
$\Sigma^{-}$	&	$0.0243 \pm 0.0005$	&$0.0233 \pm 0.0002$	&$0.0238 \pm 0.0004$ \\
$\Sigma^{0}$	&	$0.0357 \pm 0.0008$	&$0.0266 \pm 0.0003$	&$0.0287 \pm 0.0005$\\         
\noalign{\smallskip}\hline
\end{tabular}
\end{center}
\end{table*}

\begin{table*}
\begin{center}
\caption[Charged particles and hadron multiplicity mean values: \texttt{CLPOW} parameter variation effects on SHM predictions for \mbox{${\rm e}^+{\rm e}^- \rightarrow {\rm d} \bar{{\rm d}}$} collisions at $91.2$ GeV center of mass energy.]{Charged particles and hadron multiplicity mean values: \texttt{CLPOW} parameter variation effects on SHM predictions for \mbox{${\rm e}^+{\rm e}^- \rightarrow {\rm d} \bar{{\rm d}}$} collisions at $91.2$ GeV center of mass energy.}
\label{tab:multClpow}
\begin{tabular}{c c c c}
\hline\noalign{\smallskip}
 & Default & \texttt{CLPOW} = 1.00 & \texttt{CLPOW} = 4.00 \\ 
\noalign{\smallskip}\hline\noalign{\smallskip}
Charged		&	$20.29 \pm 0.32$		&$20.07 \pm 0.34$		&$20.00 \pm 0.12$\\		
$\pi^{0}$		&	$10.09 \pm 0.08$		&$10.05 \pm 0.14$		&$10.04 \pm 0.07$\\		
$\pi^{+}$		&	$8.71 \pm 0.16$		&$8.61 \pm 0.15$   		&$8.55 \pm 0.05$\\		
$\eta$		&	$1.24 \pm 0.03$		&$1.28 \pm 0.05$		&$1.22 \pm 0.02$\\		
$\Delta^{++}$	&	$0.081 \pm 0.002$		&$0.076 \pm 0.002$		&$0.0864 \pm 0.0009$\\		
$\Sigma^{+}$	&	$0.0277 \pm 0.0006$	&$0.0233 \pm 0.0005$	&$0.0239 \pm 0.0003$\\ 
$\Sigma^{-}$	&	$0.0243 \pm 0.0005$	&$0.0228 \pm 0.0005$	&$0.0246 \pm 0.0003$\\
$\Sigma^{0}$	&	$0.0357 \pm 0.0008$	&$0.0284 \pm 0.0006$	&$0.0270 \pm 0.0003$\\       
\noalign{\smallskip}\hline
\end{tabular}
\end{center}
\end{table*}

\begin{table*}
\begin{center}
\caption[Charged particles and hadron multiplicity mean values: \texttt{PSPLT} parameter variation effects on SHM predictions for \mbox{${\rm e}^+{\rm e}^- \rightarrow {\rm d} \bar{{\rm d}}$} collisions at $91.2$ GeV center of mass energy.]{Charged particles and hadron multiplicity mean values: \texttt{PSPLT} parameter variation effects on SHM predictions for \mbox{${\rm e}^+{\rm e}^- \rightarrow {\rm d} \bar{{\rm d}}$} collisions at $91.2$ GeV center of mass energy.}
\label{tab:multPsplt}
\begin{tabular}{c c c c}
\hline\noalign{\smallskip}
 & Default & \texttt{PSPLT} = 0.50 & \texttt{PSPLT} = 2.00 \\ 
\noalign{\smallskip}\hline\noalign{\smallskip}
Charged		&	$20.29 \pm 0.32$		&$18.5 \pm 0.20$		&$23.1 \pm 1.1$\\		
$\pi^{0}$		&	$10.09 \pm 0.08$		&$9.26 \pm 0.07$		&$11.6 \pm 0.5$\\		
$\pi^{+}$		&	$8.71 \pm 0.16$		&$7.98 \pm 0.10$   		&$9.9 \pm 0.5$\\		
$\eta$		&	$1.24 \pm 0.03$		&$1.14 \pm 0.03$		&$1.5 \pm 0.2$\\		
$\Delta^{++}$	&	$0.081 \pm 0.002$		&$0.067 \pm 0.001$		&$0.075 \pm 0.003$\\		
$\Sigma^{+}$	&	$0.0277 \pm 0.0006$	&$0.0205 \pm 0.0004$	&$0.029 \pm 0.001$\\ 
$\Sigma^{-}$	&	$0.0243 \pm 0.0005$	&$0.0203 \pm 0.0004$	&$0.069 \pm 0.003$\\
$\Sigma^{0}$	&	$0.0357 \pm 0.0008$	&$0.0247 \pm 0.0005$	&$0.027 \pm 0.001$\\        
\noalign{\smallskip}\hline
\end{tabular}
\end{center}
\end{table*}

\subsection{Adjustment of the free parameters}

The free parameters of the SHM code, namely $\rho$ and $\gamma_{\textnormal{S}}$, and the cluster 
minimum mass parameter $M_{\textnormal{C}}$ introduced by the merging procedure, need to be rigorously 
tuned to correctly evaluate the performances of the code in reproducing the experimental data. 
For the present work only preliminary tests have been performed, using different parameter 
configurations, in order to obtain a global overview of the SHM predictivity level in a full collision 
simulation framework. Moreover, an approximate adjustment of the SHM code free parameters has been realized, 
through a comparison between the SHM predictions and the corresponding experimental data: this comparison 
has been performed considering a set of inclusive and exclusive observable distributions and mean values. 
In particular, event shape and single particle momentum/energy distributions, single hadron mean multiplicities 
and the charged particle number distribution have been included in the analysis. The reference experimental 
data used for the comparison come from the measurements of LEP experiments for a center of mass energy 
of $91.2$ GeV. The best parameter estimation has been realized by means of an approximate $\chi^{2}$ minimization 
procedure: a set of explorative runs with various free parameter configurations has been considered, evaluating 
for each configuration the discrepancy between theoretical predictions and experimental data in the form of a 
global $\chi^{2}$ value. As usual, the parameter best configuration has been identified as the one giving 
the lowest global $\chi^{2}$ value. More in detail, for histograms the $\chi^{2}$ value has been computed as the sum 
over channels of the discrepancy between the theoretical prediction $y^{i}_{t}$ and the corresponding 
experimental data $y^{i}_{e}$, normalized using the theoretical (Monte-Carlo) and experimental errors 
$\sigma^{i}_{t}$ and $\sigma^{i}_{e}$:

\begin{equation}
\chi^{2} = \sum_{i=1}^{N}\frac{\left(y^{i}_{t}-y^{i}_{e}\right)^{2}}{\sigma^{i2}_{t}+\sigma^{i2}_{e}},
\end{equation}
where $i$ is the histogram channel index and $N$ the number of channels. 
The $\chi^{2}$ value corresponding to the particle multiplicity analysis has been computed in a similar way, 
summing over the list of considered particles. Finally, the global $\chi^{2}$ value has been computed as the 
sum over the single distribution/mean value $\chi^{2}$ contributions.

The performed analysis, whose details are reported in \cite{phdthesis}, shows that the parameter setup 
corresponding to the best global agreement between SHM predictions and the corresponding 
experimental data is:
\\
\\
\begin{itemize}
\item $M_{\textnormal{C}} = 1.6$ GeV;
\item $\rho = 0.45$  GeV/$\textnormal{fm}^{3}$;
\item $\gamma_{\textnormal{S}} = 0.65$.
\end{itemize} 

\section{Numerical results}
\label{sec:NumericalResults}

In this Section, we present the results obtained with the SHM code for the above parameter 
set. The SHM predictions are compared to the corresponding predictions of 
\texttt{Herwig6510} in its release version and to experimental data. As already mentioned, the reported 
results refer to \ee collisions at the Z peak. It must be noted that the chosen observables provide 
a reliable measure for the evaluation of the hadronization model predictivity, because of their sensitivity 
to the hadronization process. This condition is clear for what concerns exclusive observables 
involving a single hadron species, such as the $\pi^{\pm}$ moment distribution and mean multiplicity. 
Nevertheless, the same property holds also for the inclusive event shape observables referring to the 
event transverse plane, such as transverse momentum: the main contribution to the global hadron 
distribution in the event phase space comes from the hard scattering, whose quark emission direction 
appro\-ximatively defines the event thrust axis. On the other side, the hadronic phase space configuration 
projection on the event transverse plane is strongly influenced by the hadronization process. Therefore, 
the analysis of the transverse plane related observables allows to effectively evaluate the prediction 
accuracy of the considered hadronization model. The results obtained for a subset of these inclusive 
and exclusive observables are reported hereafter (for the definition of observables, see Appendix).

\begin{figure}
\includegraphics[width=0.450\textwidth]{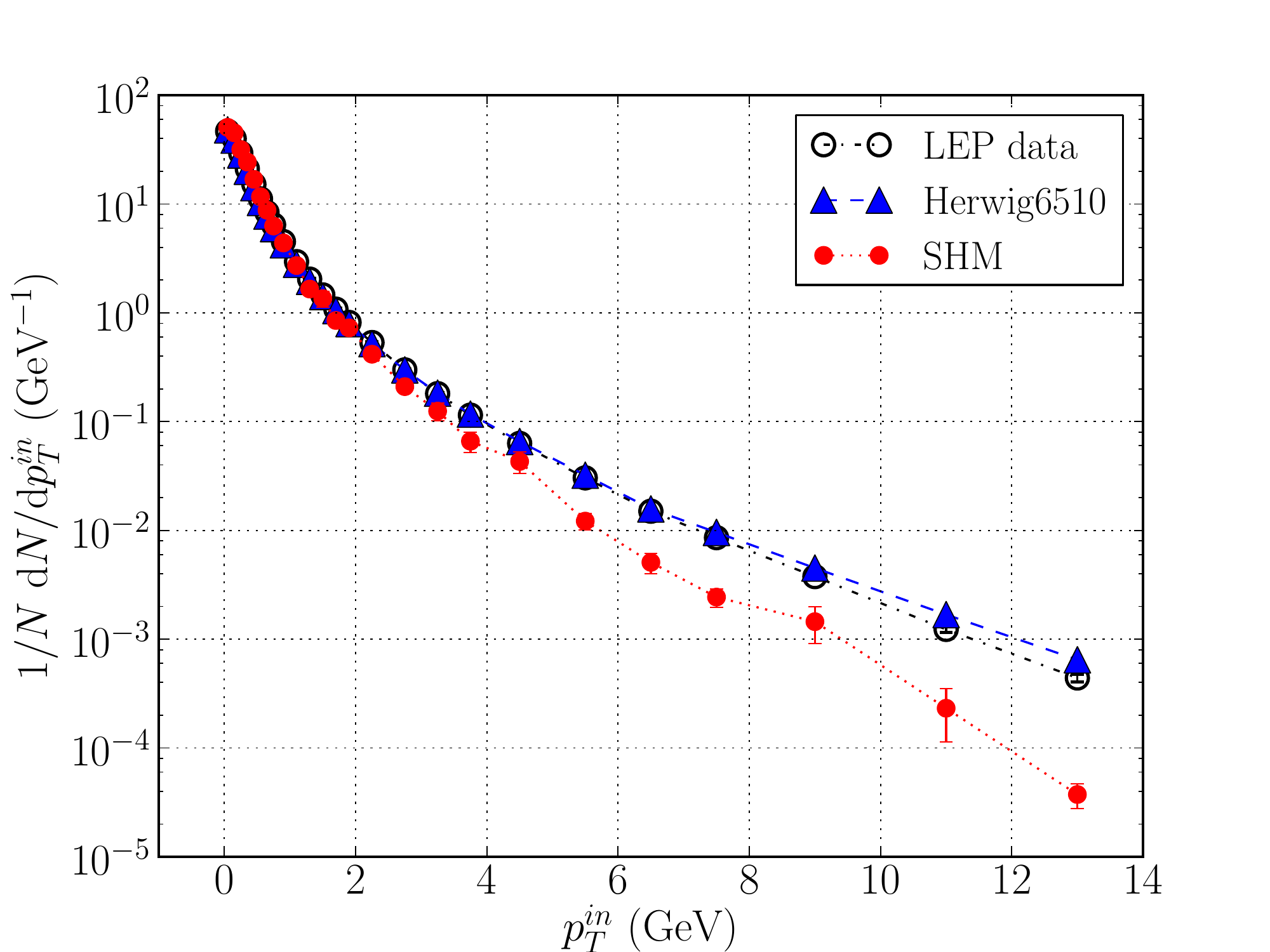}
\caption[Transverse momentum (in) with respect to thrust axis normalized distribution: comparison among SHM predictions, \mbox{\texttt{Herwig6510}} predictions and DELPHI data \cite{DELPHI2}.]{Transverse momentum (in) with respect to thrust axis normalized distribution: comparison among SHM predictions, \mbox{\texttt{Herwig6510}} predictions and DELPHI data \cite{DELPHI2}.}
\label{fig:ptT_in}
\end{figure}

\begin{figure}
\includegraphics[width=0.450\textwidth]{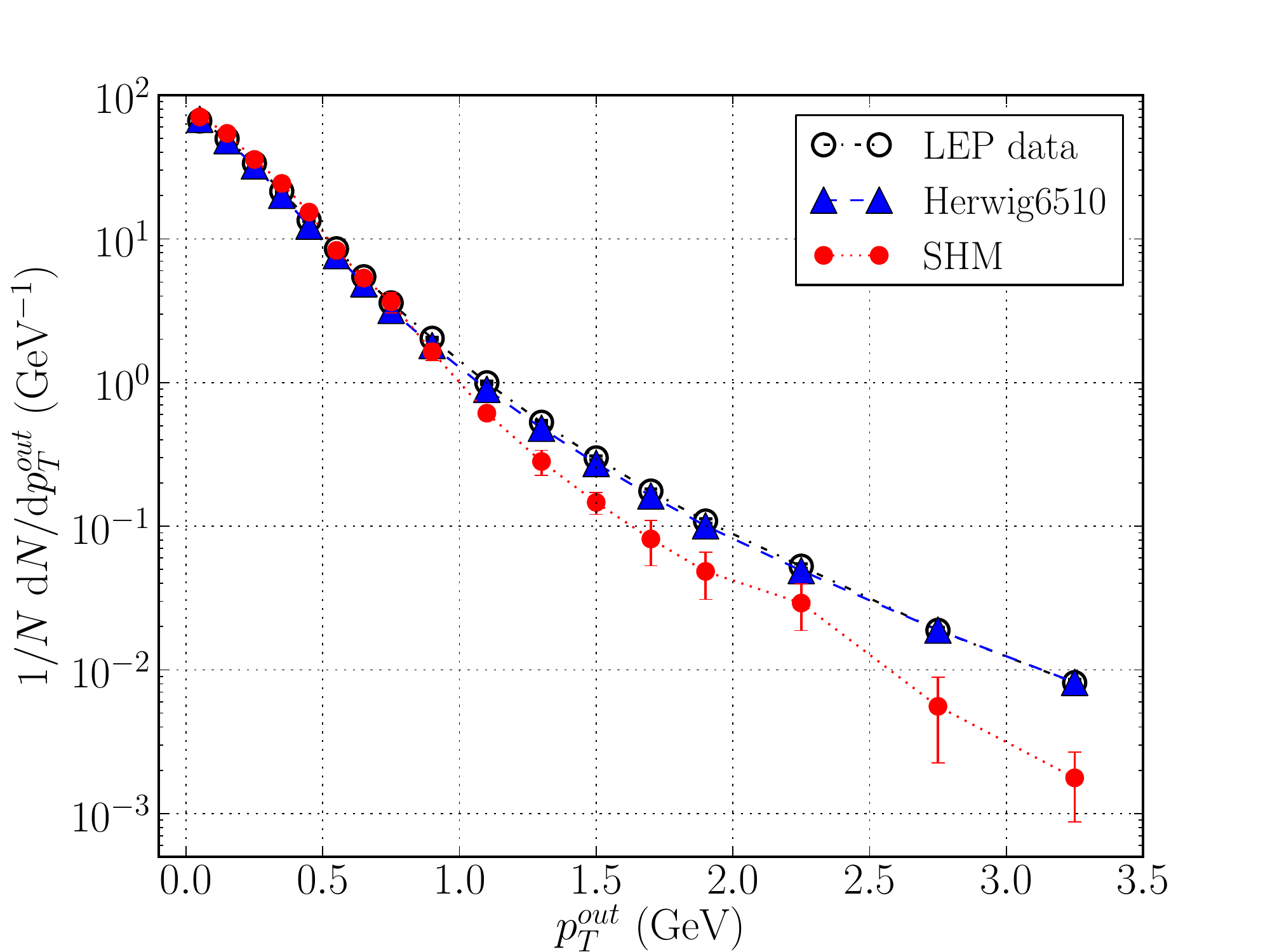}
\caption[Transverse momentum (out) with respect to thrust axis normalized distribution: comparison among SHM predictions, \mbox{\texttt{Herwig6510}} predictions and DELPHI data \cite{DELPHI2}.]{Transverse momentum (out) with respect to thrust axis normalized distribution: comparison among SHM predictions, \mbox{\texttt{Herwig6510}} predictions and DELPHI data \cite{DELPHI2}.}
\label{fig:ptT_out}
\end{figure}

\begin{figure}
\includegraphics[width=0.450\textwidth]{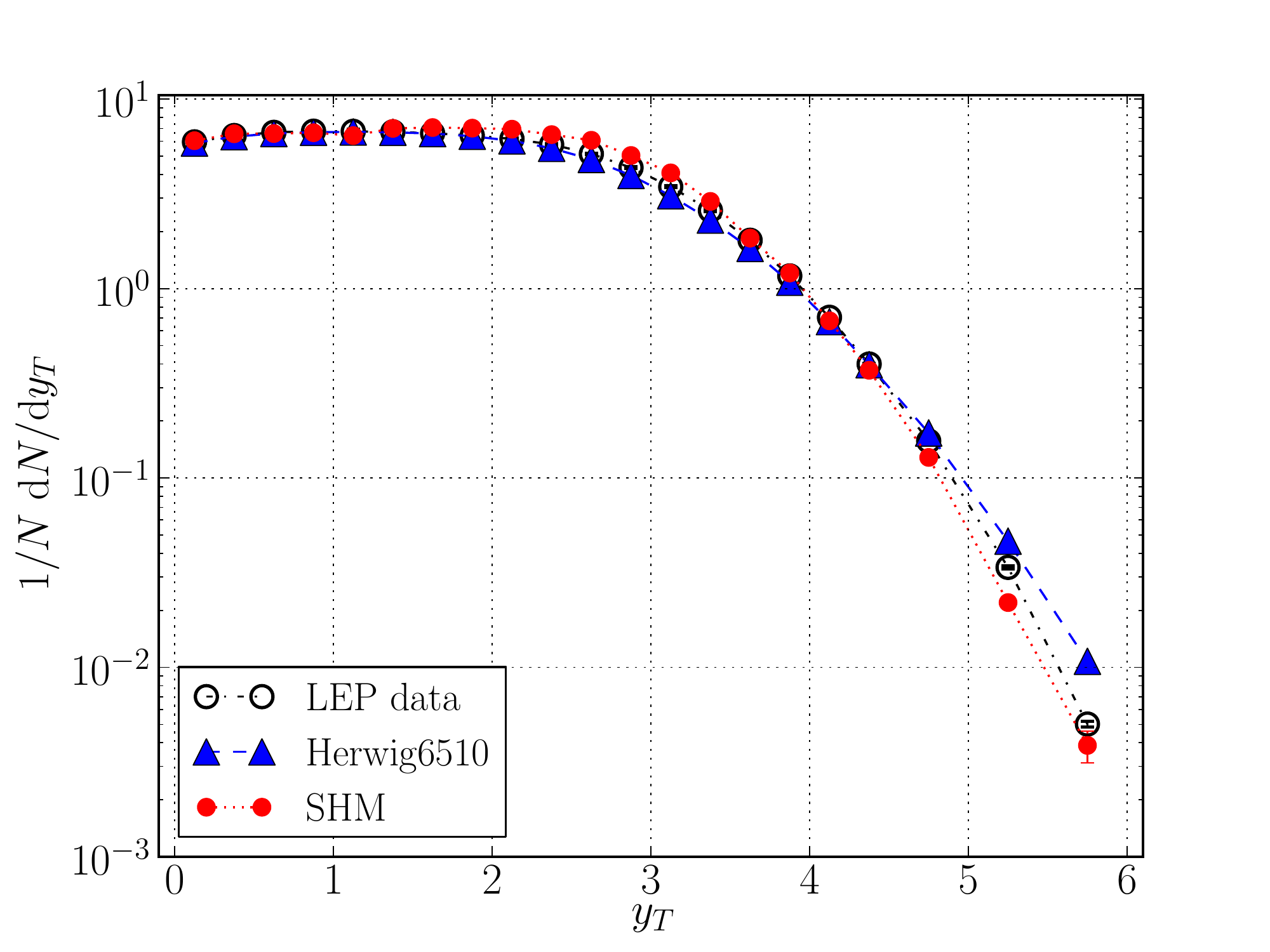}
\caption[Rapidity with respect to thrust axis normalized distribution: comparison among SHM predictions, \mbox{\texttt{Herwig6510}} predictions and DELPHI data \cite{DELPHI2}.]{Rapidity with respect to thrust axis normalized distribution: comparison among SHM predictions, \mbox{\texttt{Herwig6510}} predictions and DELPHI data \cite{DELPHI2}.}
\label{fig:y_T}
\end{figure}

\begin{figure}
\includegraphics[width = 0.450\textwidth]{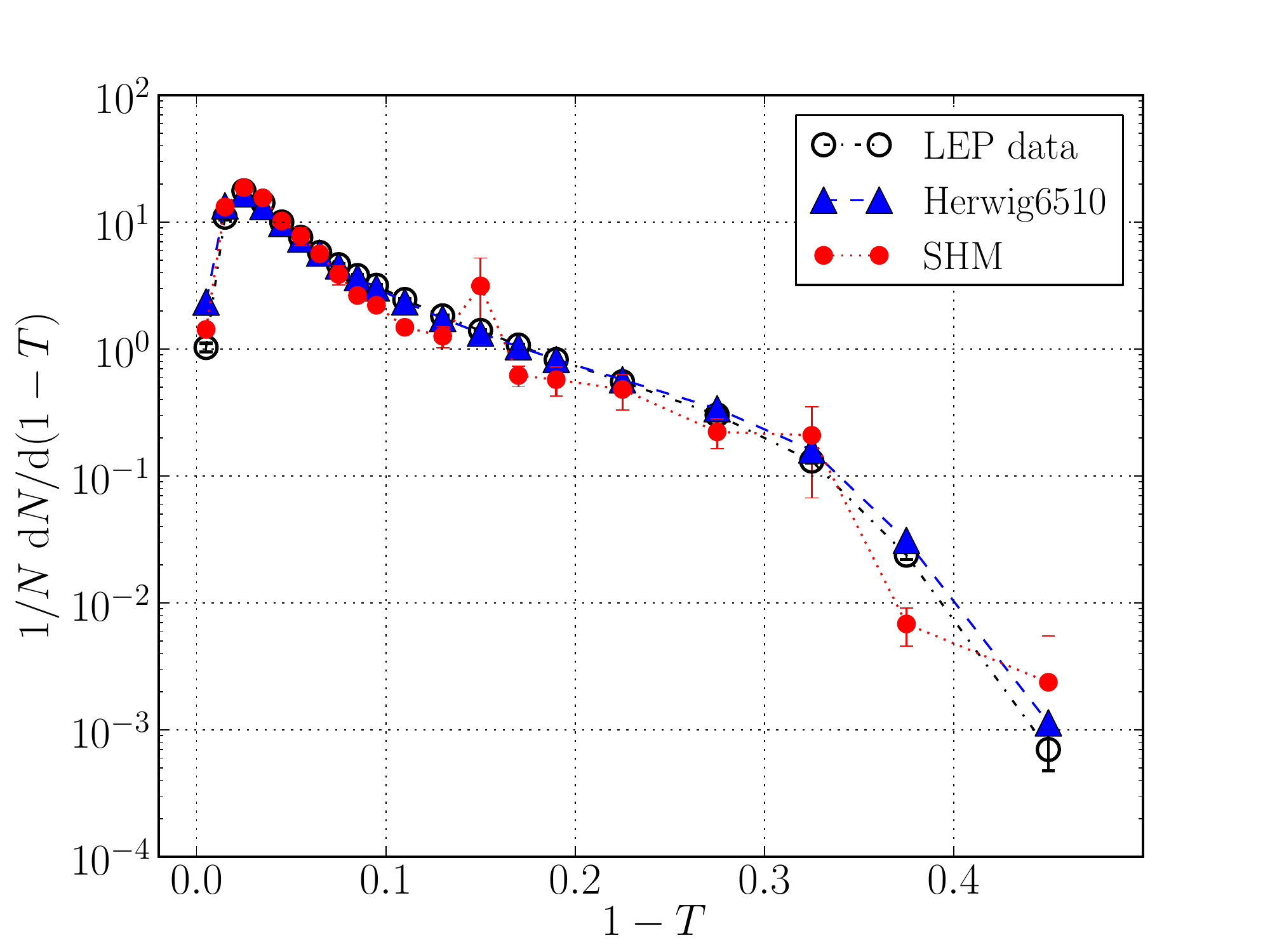}
\caption[Thrust normalized distribution: comparison among SHM predictions, \mbox{\texttt{Herwig6510}} predictions and DELPHI data \cite{DELPHI2}.]{Thrust normalized distribution: comparison among SHM predictions, \mbox{\texttt{Herwig6510}} predictions and DELPHI data \cite{DELPHI2}.}
\label{fig:1_T}
\end{figure}

\begin{figure}
\includegraphics[width=0.450\textwidth]{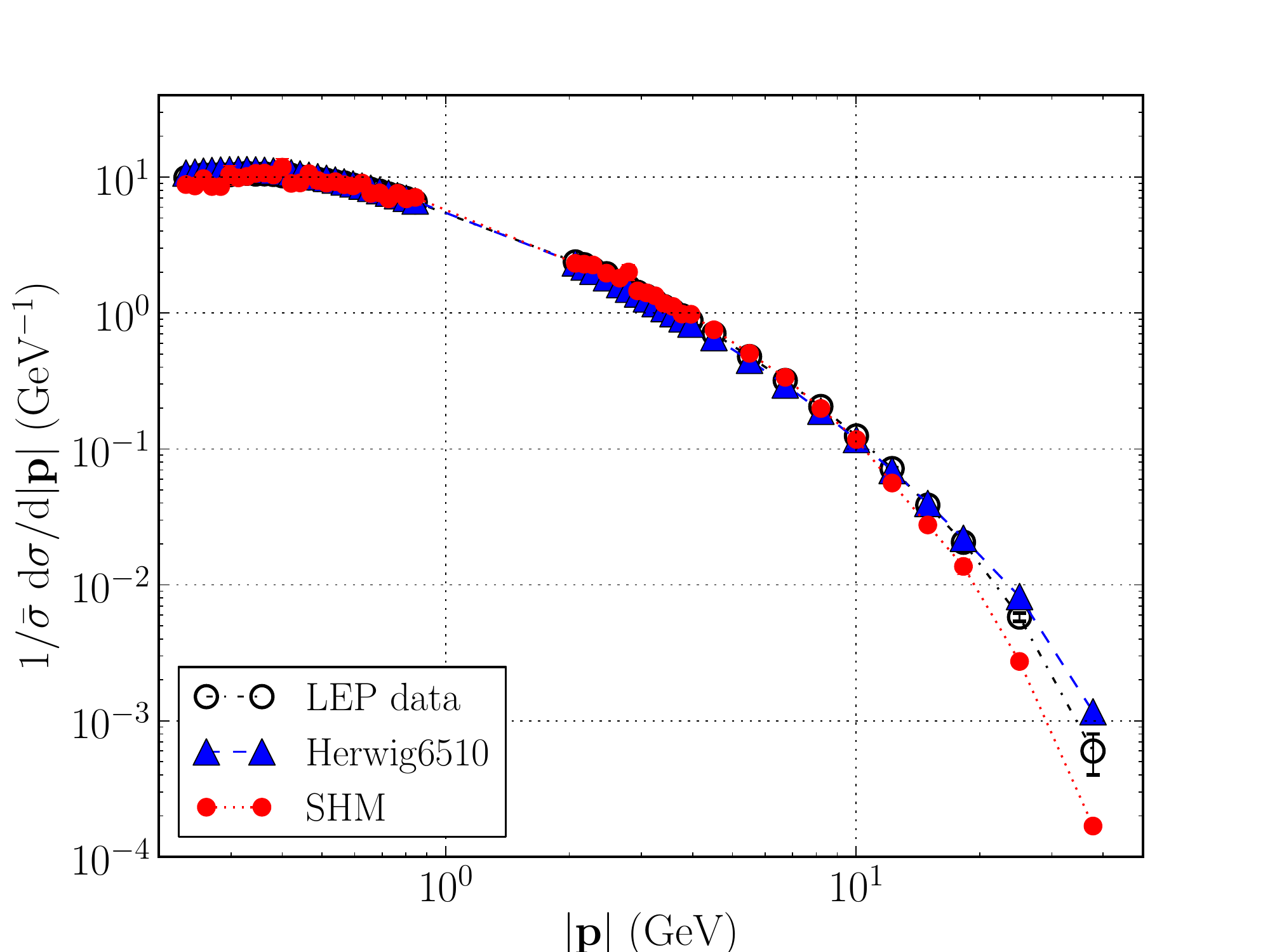}
\caption[$\pi^{\pm}$ 3-momentum module distribution: comparison among SHM predictions, \mbox{\texttt{Herwig6510}} predictions and OPAL data \cite{OPAL2}.]{$\pi^{\pm}$ 3-momentum module distribution: comparison among SHM predictions, \mbox{\texttt{Herwig6510}} predictions and OPAL data \cite{OPAL2}.}
\label{fig:p_picharged}
\end{figure}

\begin{figure}
\includegraphics[width=0.450\textwidth]{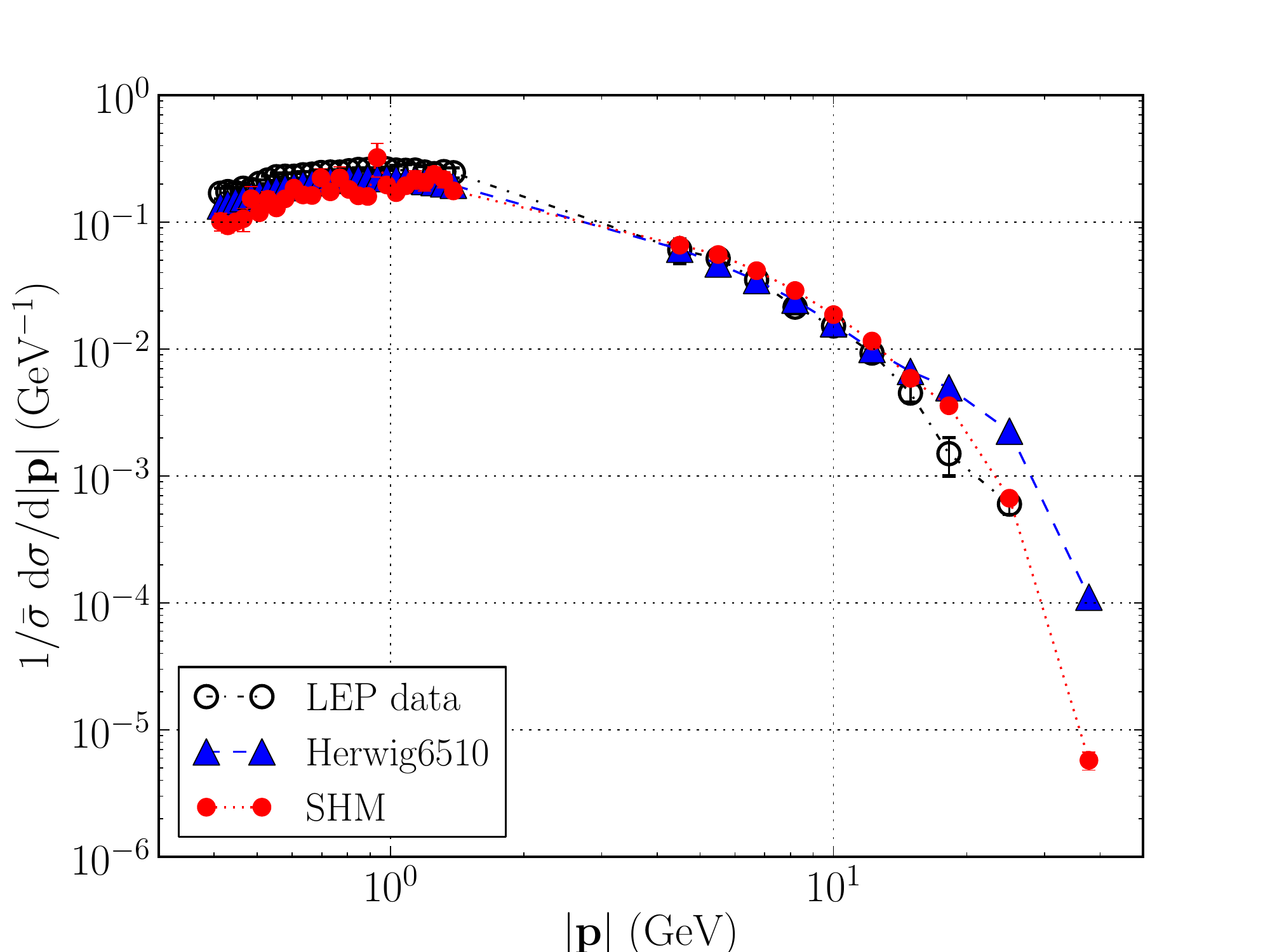}
\caption[Proton (antiproton) 3-momentum module distribution: comparison among SHM predictions, \mbox{\texttt{Herwig6510}} predictions and OPAL data \cite{OPAL2}.]{Proton (antiproton) 3-momentum module distribution: comparison among SHM predictions, \mbox{\texttt{Herwig6510}} predictions and OPAL data \cite{OPAL2}.}
\label{fig:p_proton}
\end{figure}

\begin{figure}
\includegraphics[width=0.450\textwidth]{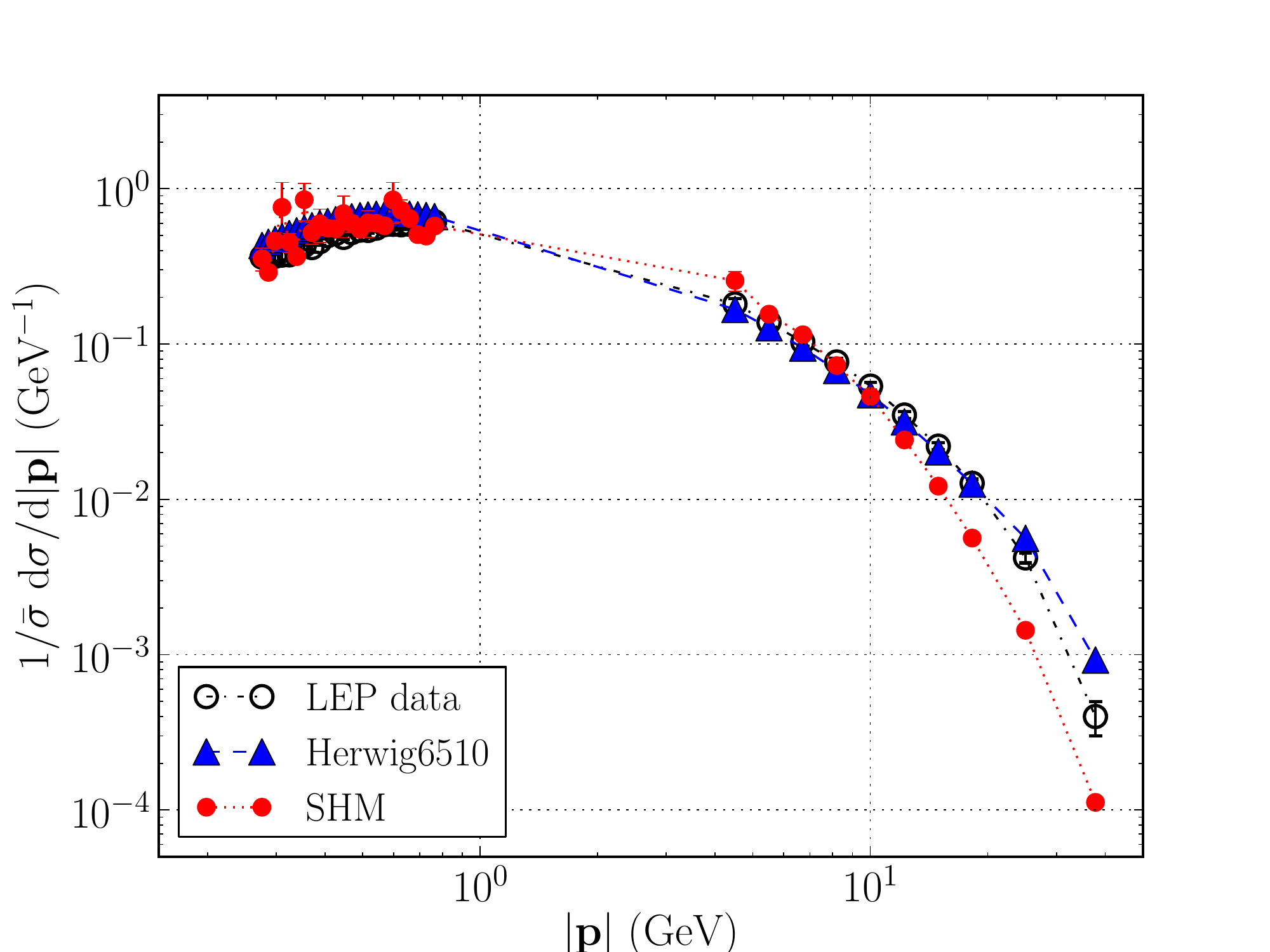}
\caption[${\rm K}^{\pm}$ 3-momentum module distribution: comparison among SHM predictions, \mbox{\texttt{Herwig6510}} predictions and OPAL data \cite{OPAL2}.]{${\rm K}^{\pm}$ 3-momentum module distribution: comparison among SHM predictions, \mbox{\texttt{Herwig6510}} predictions and OPAL data \cite{OPAL2}.}
\label{fig:p_Kcharged}
\end{figure}

\begin{figure}
\includegraphics[width=0.450\textwidth]{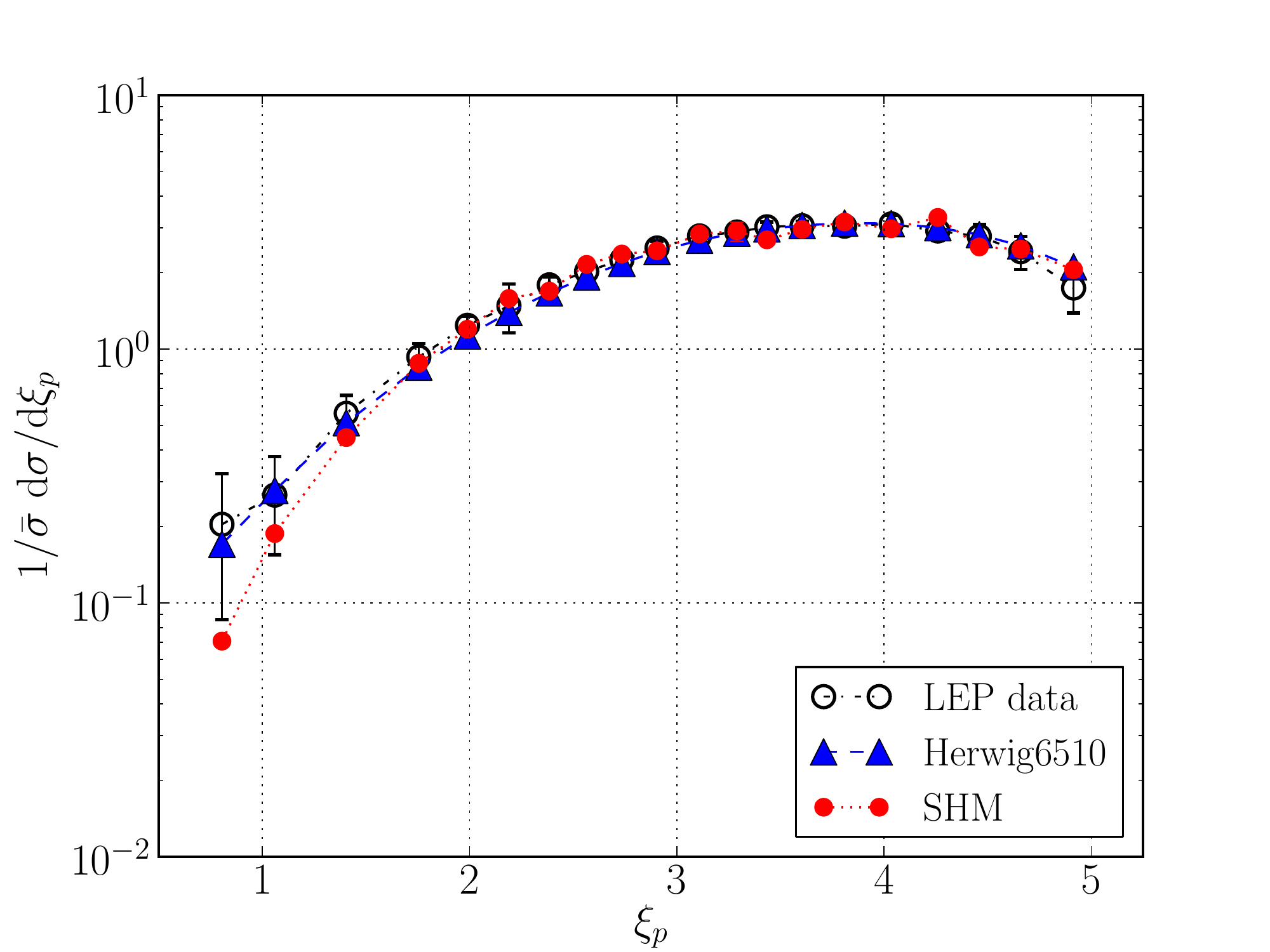}
\caption[$\pi^{0}$ scaled momentum distribution: comparison among SHM predictions, \mbox{\texttt{Herwig6510}} predictions and OPAL data \cite{OPAL8}.]{$\pi^{0}$ scaled momentum distribution: comparison among SHM predictions, \mbox{\texttt{Herwig6510}} predictions and OPAL data \cite{OPAL8}.}
\label{fig:l_x_p_pizero}
\end{figure}

\begin{figure}
\includegraphics[width=0.450\textwidth]{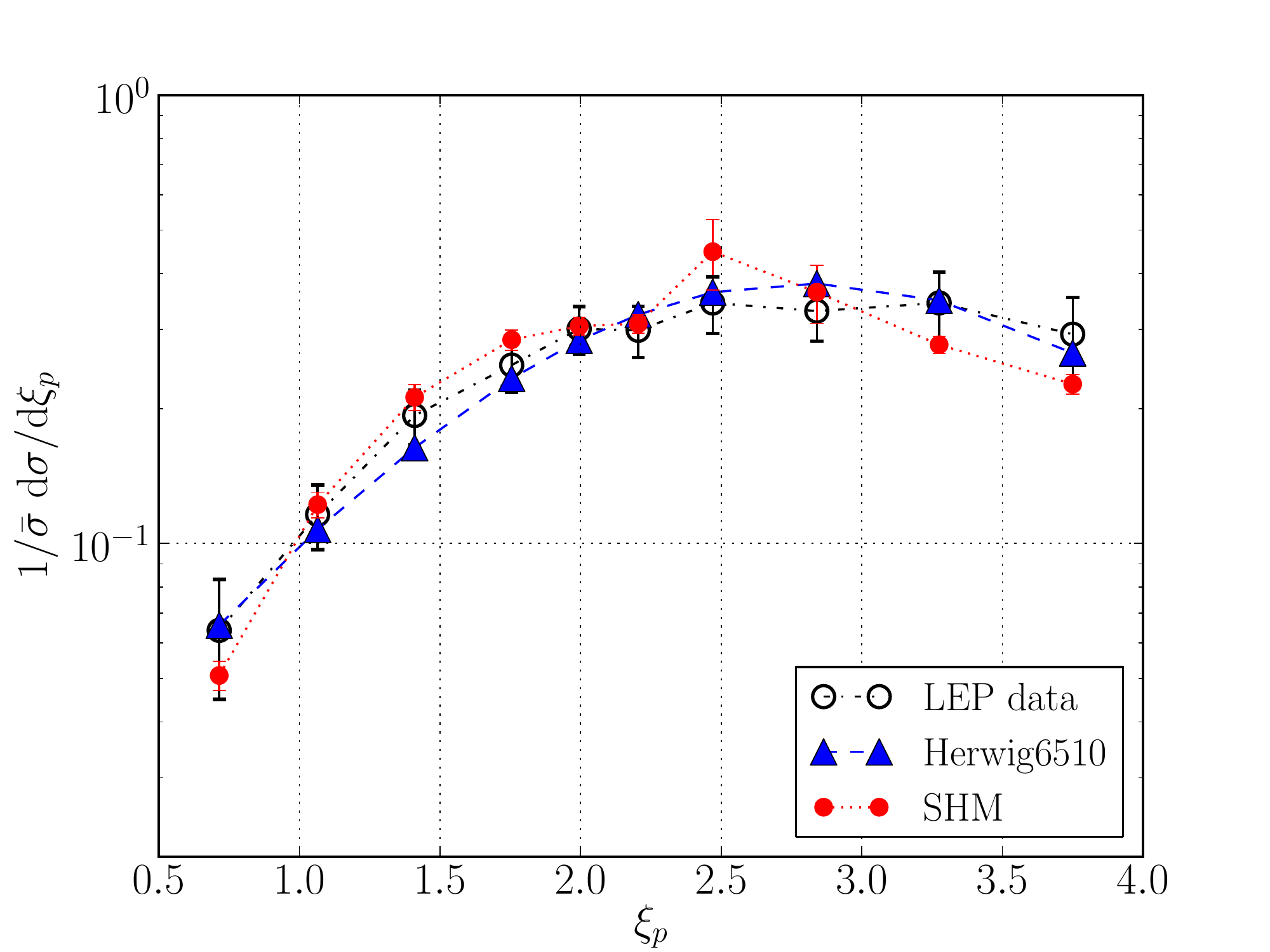}
\caption[$\omega$ scaled momentum distribution: comparison among SHM predictions, \mbox{\texttt{Herwig6510}} predictions and OPAL data \cite{OPAL8}.]{$\omega$ scaled momentum distribution: comparison among SHM predictions, \mbox{\texttt{Herwig6510}} predictions and OPAL data \cite{OPAL8}.}
\label{fig:l_x_p_omega}
\end{figure}

\begin{figure}
\includegraphics[width=0.450\textwidth]{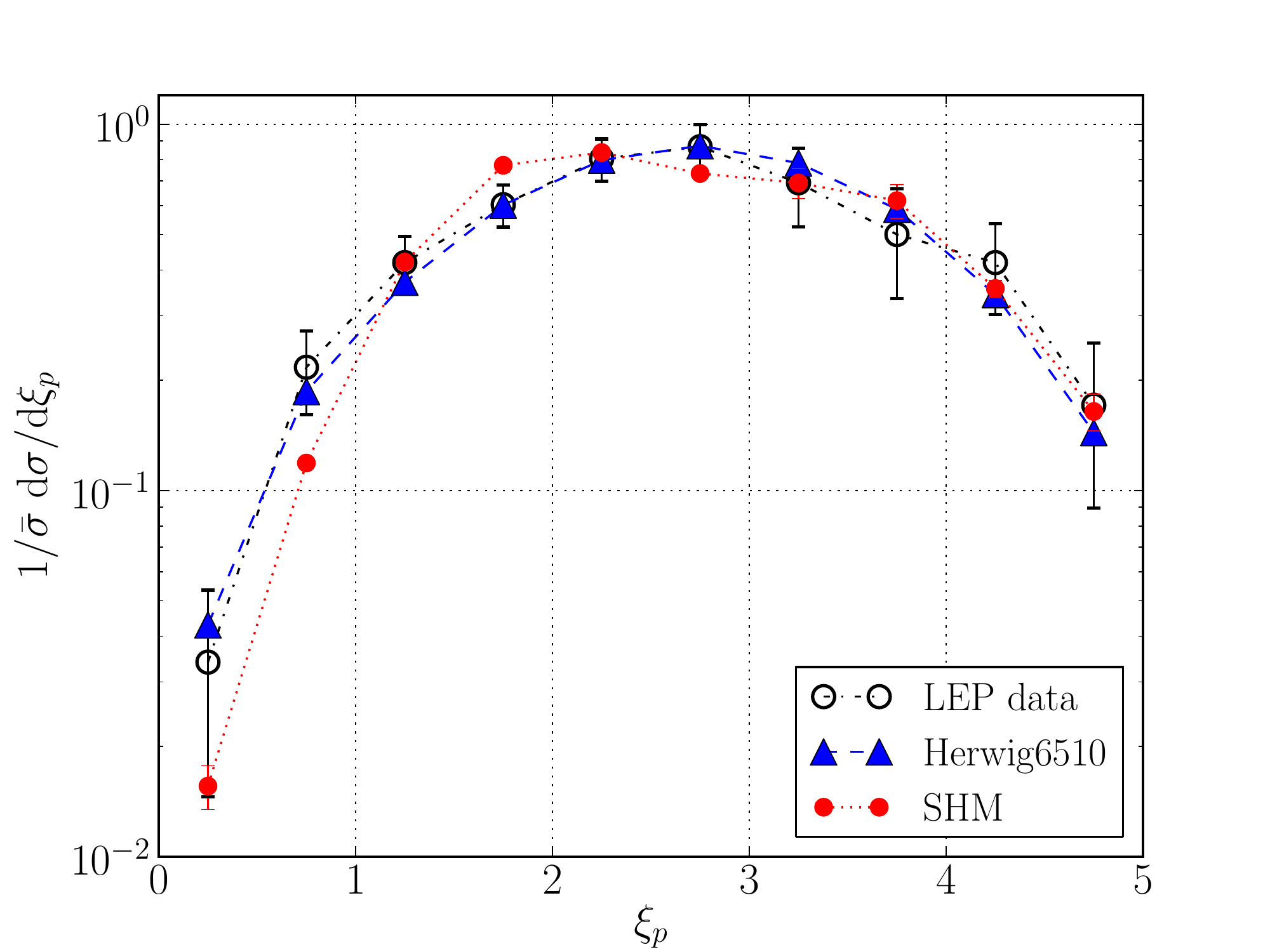}
\caption[$\rho^{\pm}$ scaled momentum distribution: comparison among SHM predictions, \mbox{\texttt{Herwig6510}} predictions and OPAL data \cite{OPAL8}.]{$\rho^{\pm}$ scaled momentum distribution: comparison among SHM predictions, \mbox{\texttt{Herwig6510}} predictions and OPAL data \cite{OPAL8}.}
\label{fig:l_x_p_rhop}
\end{figure}

\begin{figure}
\includegraphics[width=0.450\textwidth]{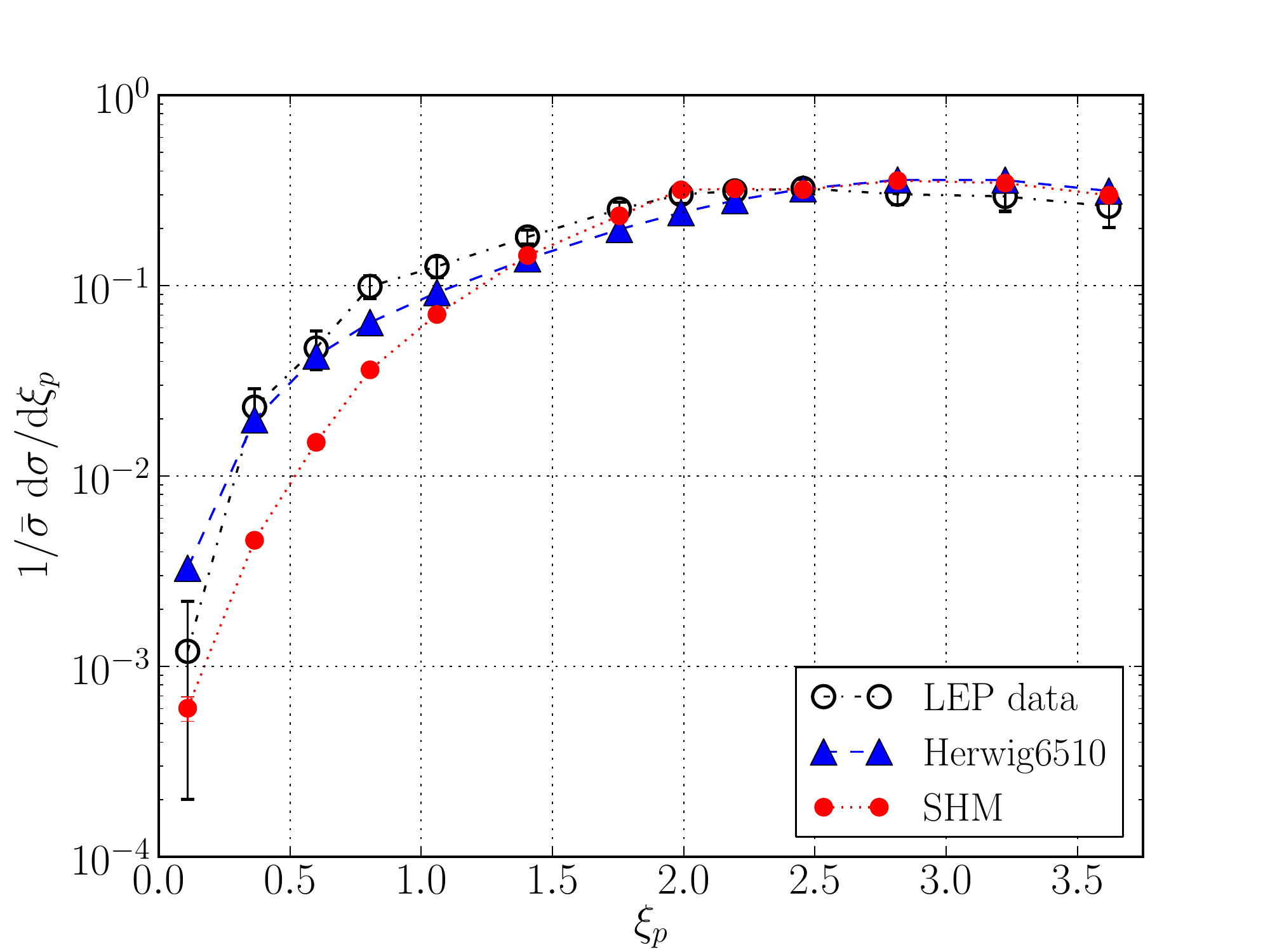}
\caption[$\eta$ scaled momentum distribution: comparison among SHM predictions, \mbox{\texttt{Herwig6510}} predictions and OPAL data \cite{OPAL8}.]{$\eta$ scaled momentum distribution: comparison among SHM predictions, \mbox{\texttt{Herwig6510}} predictions and OPAL data \cite{OPAL8}.}
\label{fig:l_x_p_eta}
\end{figure}

\begin{figure}
\includegraphics[width=0.450\textwidth]{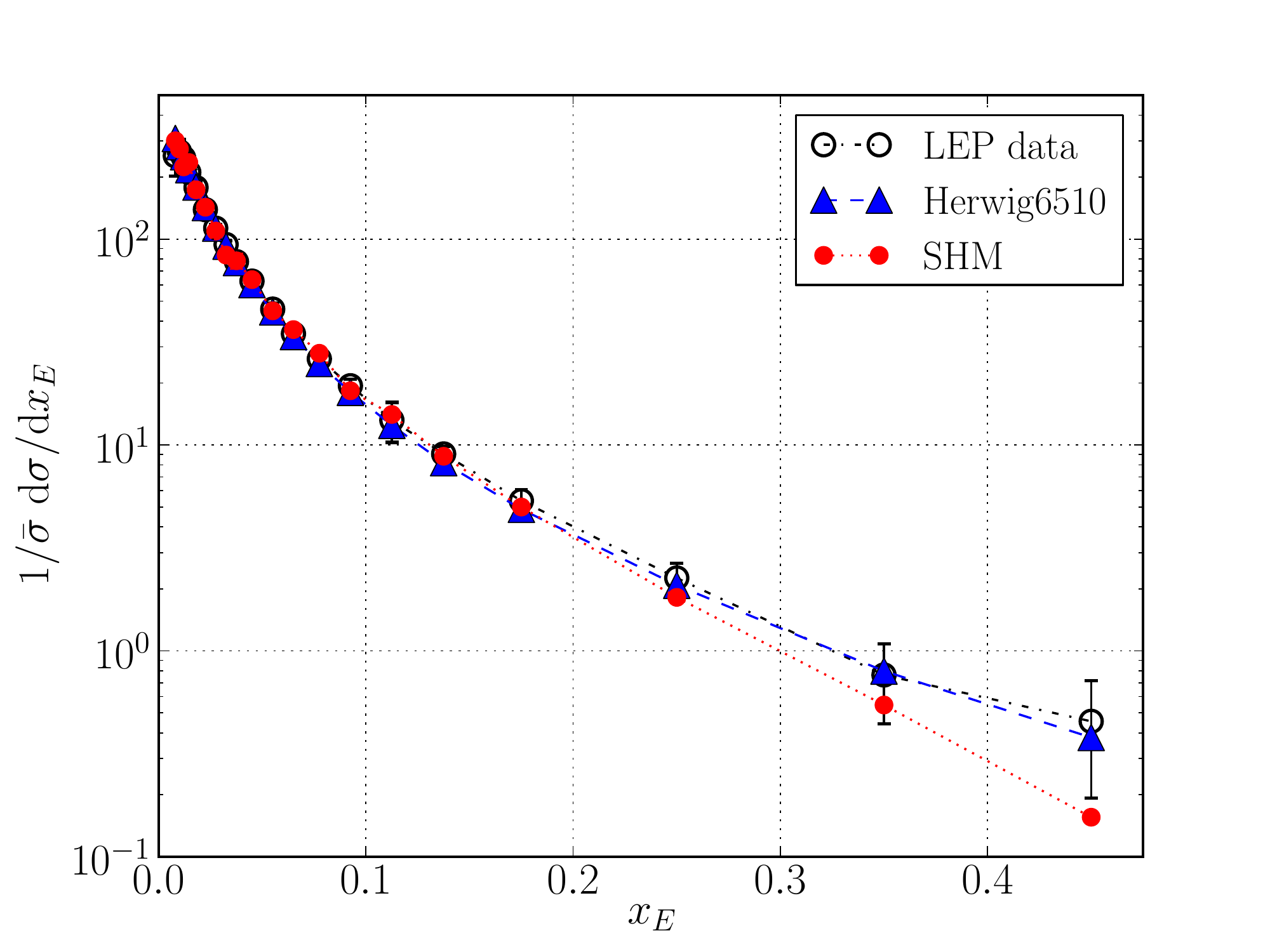}
\caption[$\pi^{0}$ scaled energy distribution: comparison among SHM predictions, \mbox{\texttt{Herwig6510}} predictions and OPAL data \cite{OPAL8}.]{$\pi^{0}$ scaled energy  distribution: comparison among SHM predictions, \mbox{\texttt{Herwig6510}} predictions and OPAL data \cite{OPAL8}.}
\label{fig:x_e_pizero}
\end{figure}

\begin{figure}
\includegraphics[width=0.450\textwidth]{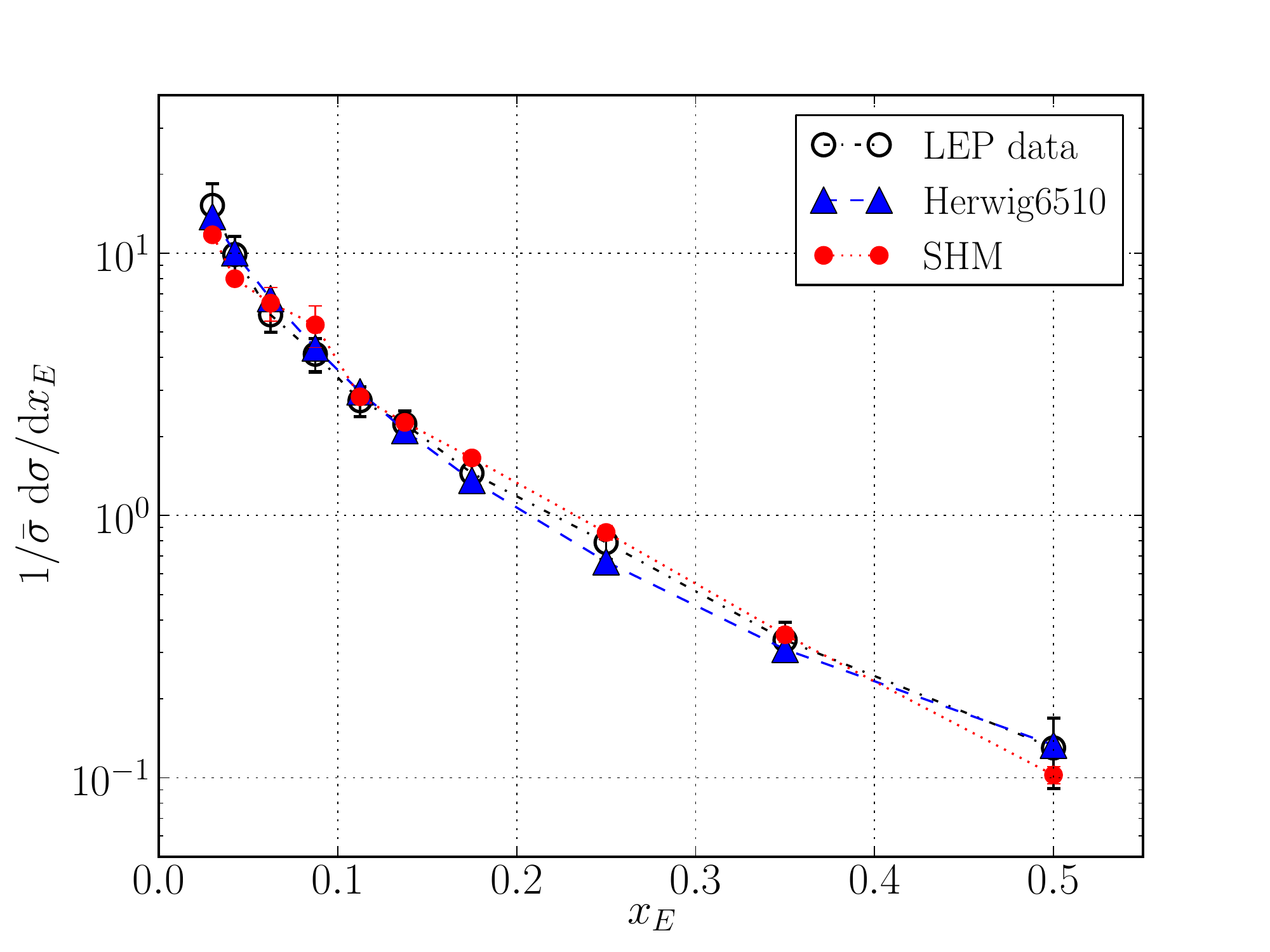}
\caption[$\omega$ scaled energy distribution: comparison among SHM predictions, \mbox{\texttt{Herwig6510}} predictions and OPAL data \cite{OPAL8}.]{$\omega$ scaled energy distribution: comparison among SHM predictions, \mbox{\texttt{Herwig6510}} predictions and OPAL data \cite{OPAL8}.}
\label{fig:x_e_omega}
\end{figure}

\begin{figure}
\includegraphics[width=0.450\textwidth]{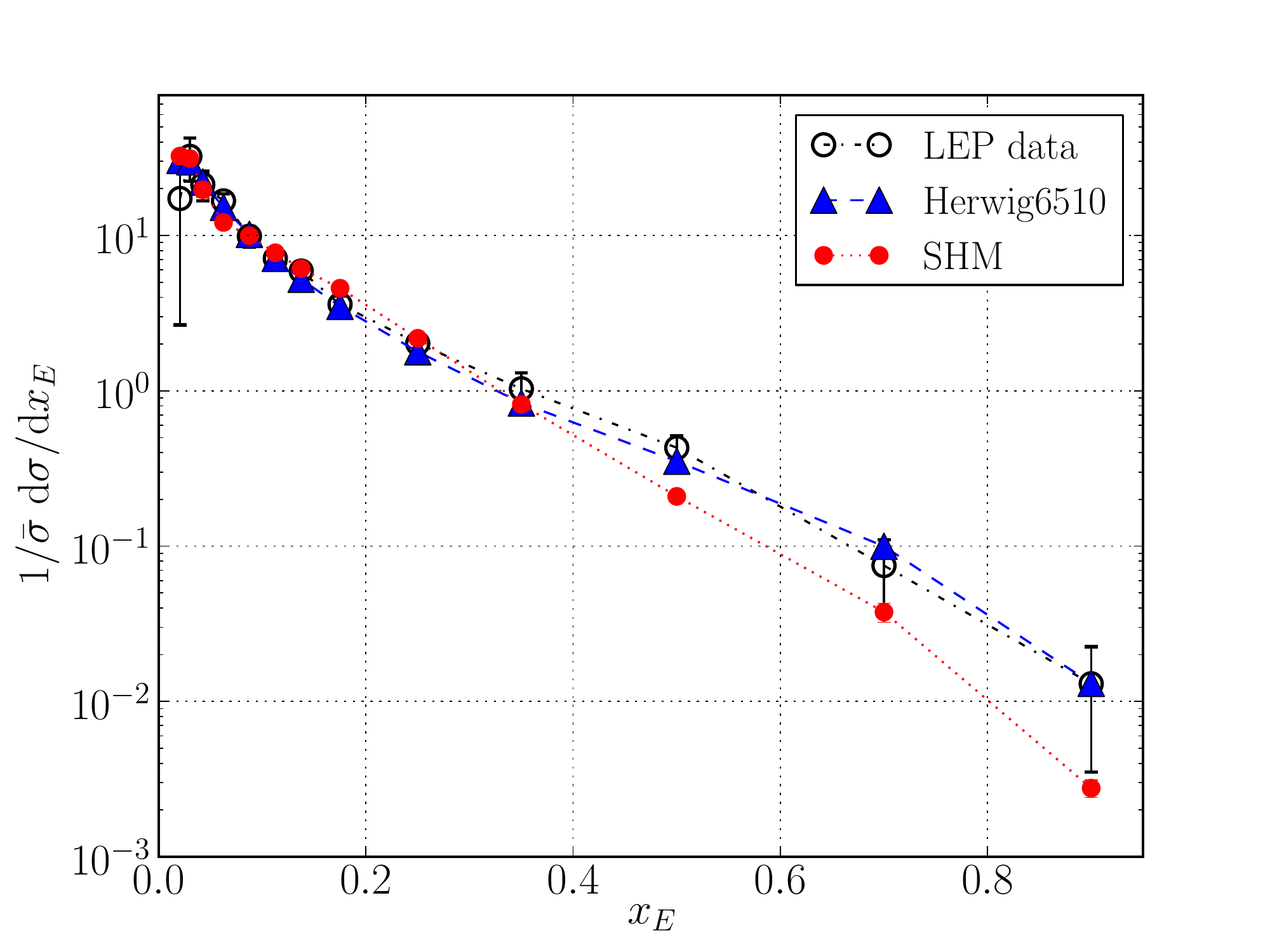}
\caption[$\rho^{\pm}$ scaled energy distribution: comparison among SHM predictions, \mbox{\texttt{Herwig6510}} predictions and OPAL data \cite{OPAL8}.]{$\rho^{\pm}$ scaled energy distribution: comparison among SHM predictions, \mbox{\texttt{Herwig6510}} predictions and OPAL data \cite{OPAL8}.}
\label{fig:x_e_rhop}
\end{figure}

\begin{figure}
\includegraphics[width=0.450\textwidth]{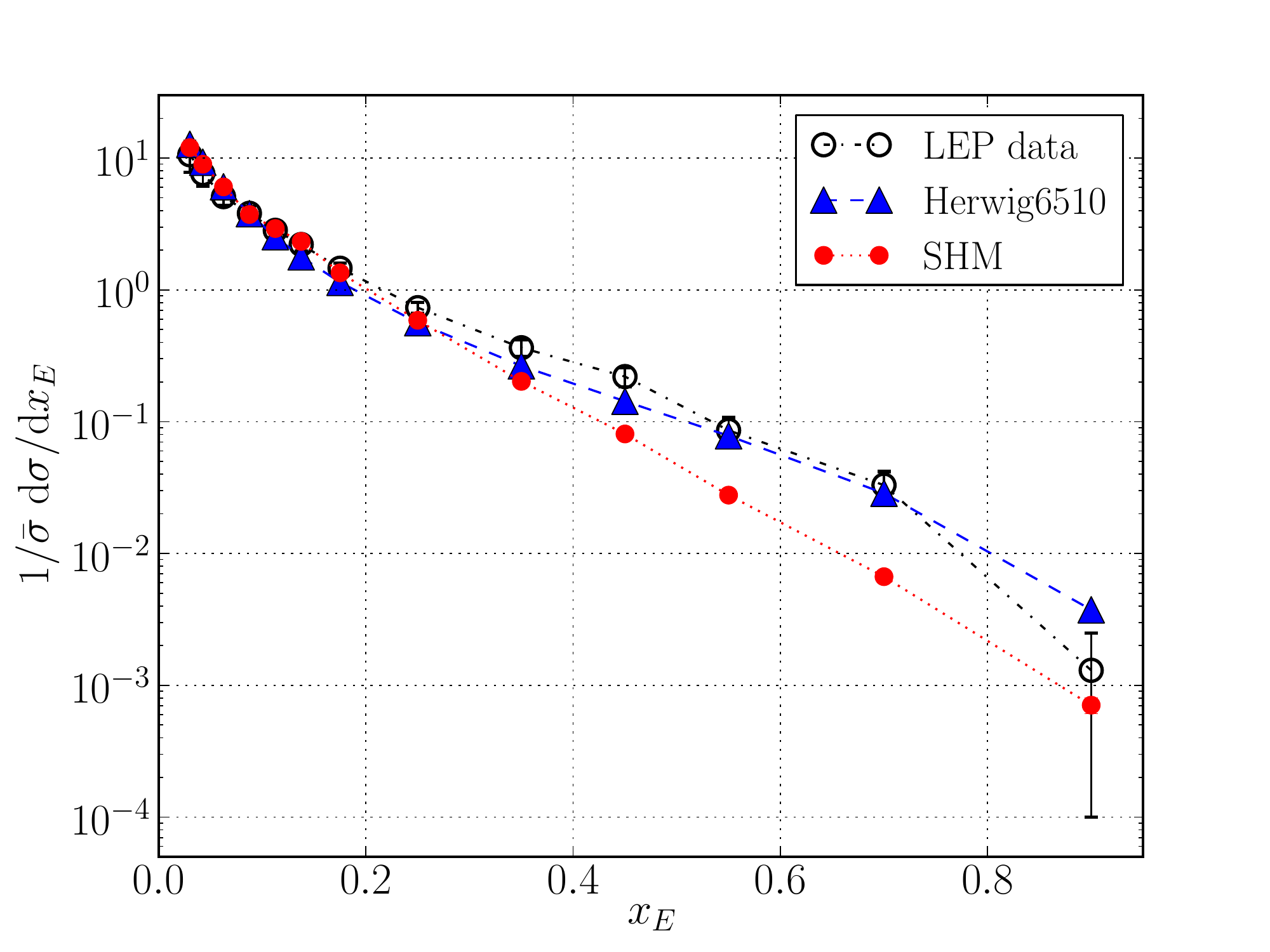}
\caption[$\eta$ scaled energy distribution: comparison among SHM predictions, \mbox{\texttt{Herwig6510}} predictions and OPAL data \cite{OPAL8}.]{$\eta$ scaled energy distribution: comparison among SHM predictions, \mbox{\texttt{Herwig6510}} predictions and OPAL data \cite{OPAL8}.}
\label{fig:x_e_eta}
\end{figure}

\begin{figure}
\includegraphics[width=0.450\textwidth]{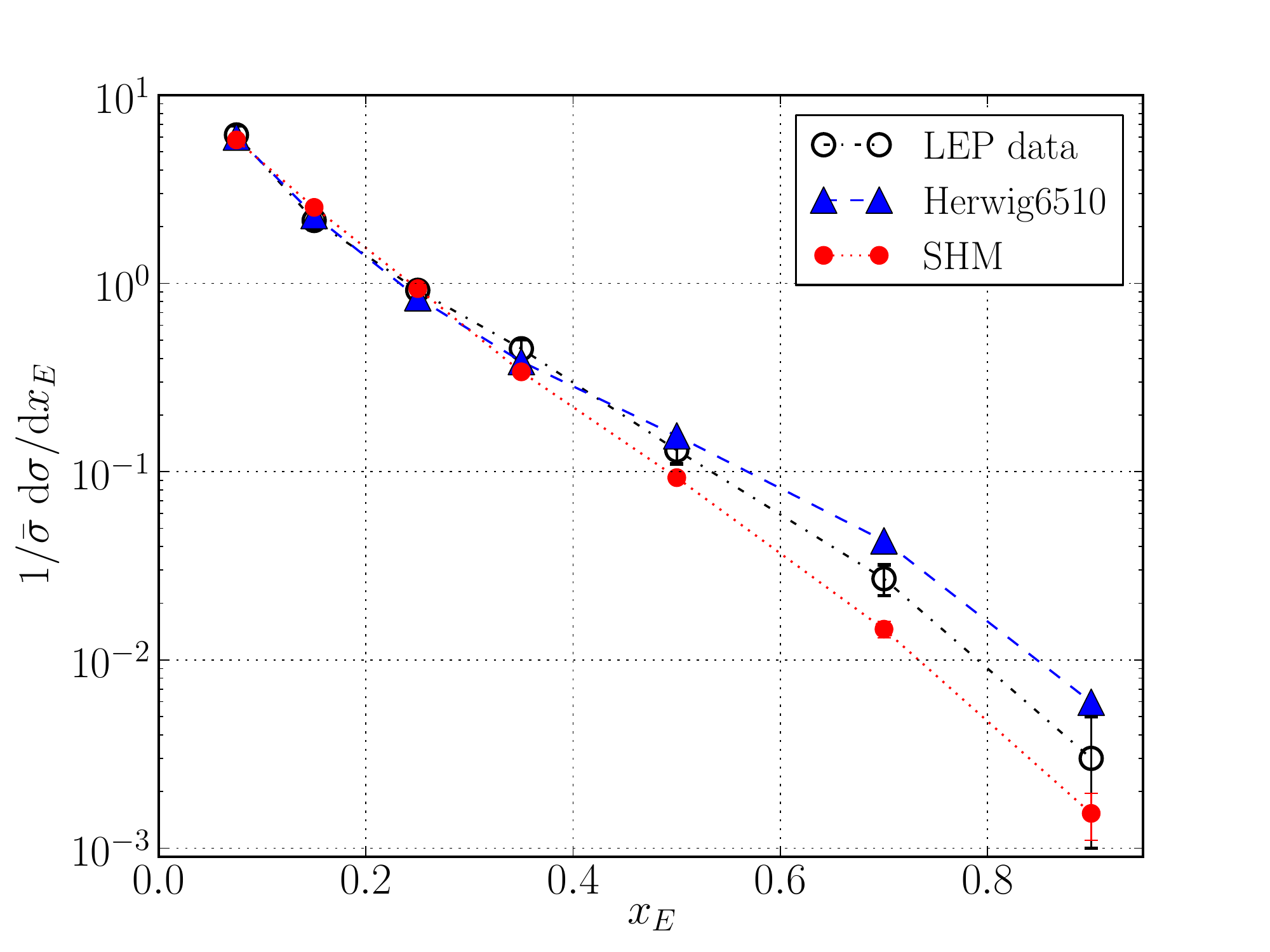}
\caption[$\rho^{0}$ scaled energy distribution: comparison among SHM predictions, \mbox{\texttt{Herwig6510}} predictions and DELPHI data \cite{DELPHI3}.]{$\rho^{0}$ scaled energy distribution: comparison among SHM predictions, \mbox{\texttt{Herwig6510}} predictions and DELPHI data \cite{DELPHI3}.}
\label{fig:x_e_rhozero}
\end{figure}

\begin{figure}
\includegraphics[width=0.450\textwidth]{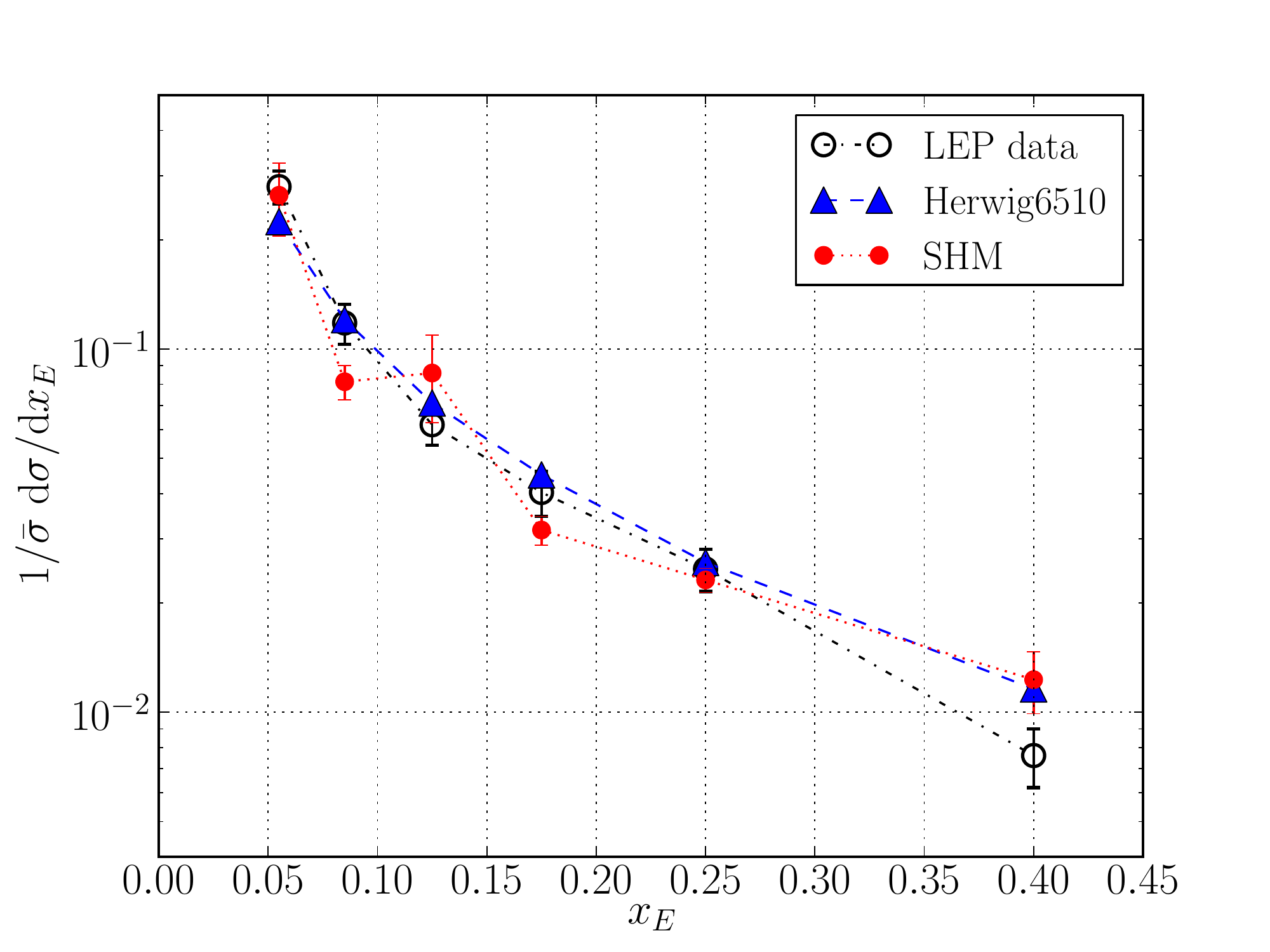}
\caption[$\Sigma^{*+}$ scaled energy distribution: comparison among SHM predictions, \mbox{\texttt{Herwig6510}} predictions and OPAL data \cite{OPAL4}.]{$\Sigma^{*+}$ scaled energy distribution: comparison among SHM predictions, \mbox{\texttt{Herwig6510}} predictions and OPAL data \cite{OPAL4}.}
\label{fig:x_e_sigmastp}
\end{figure}

\begin{figure}
\includegraphics[width=0.450\textwidth]{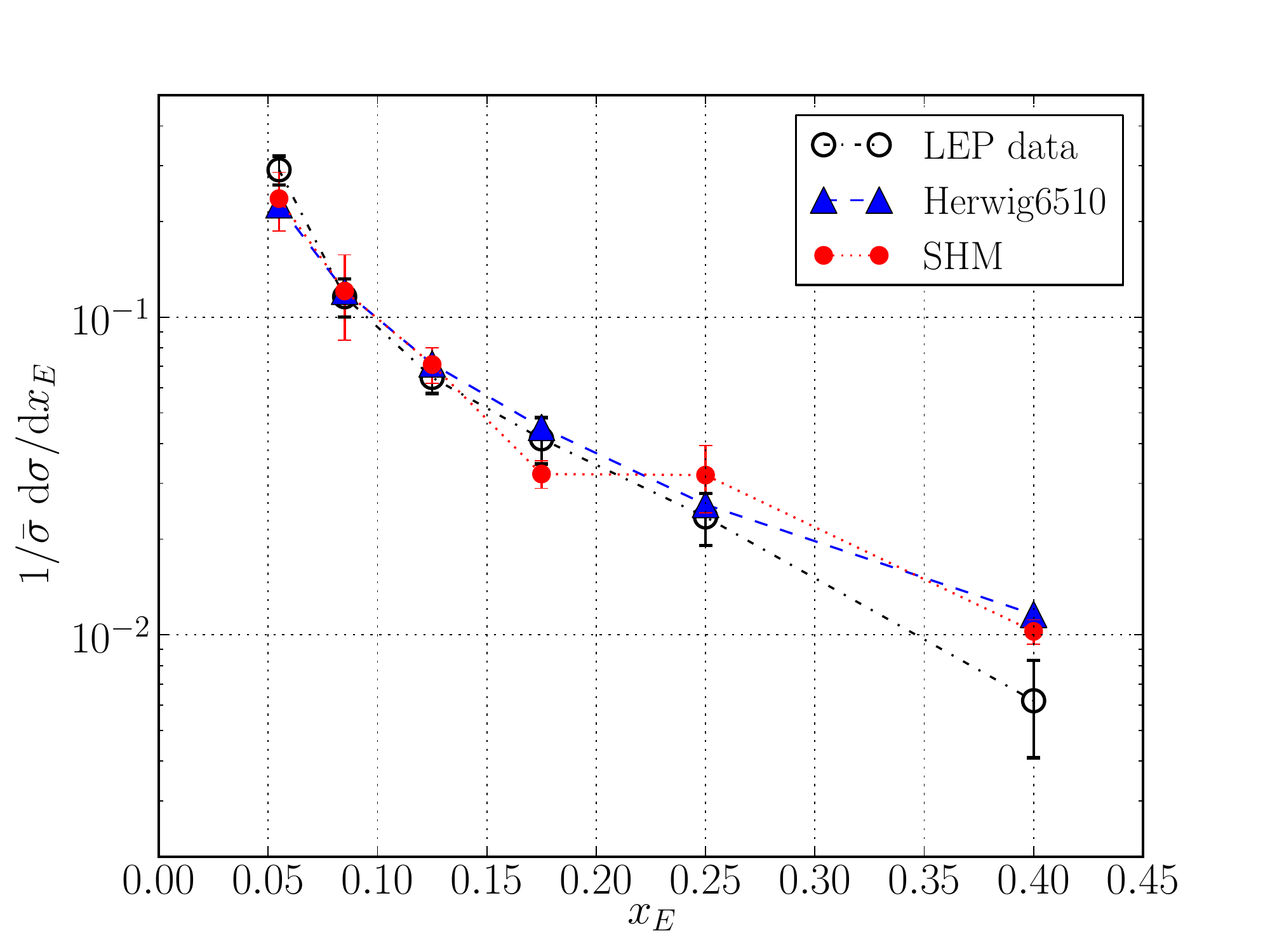}
\caption[$\Sigma^{*-}$ scaled energy distribution: comparison among SHM predictions, \mbox{\texttt{Herwig6510}} predictions and OPAL data \cite{OPAL4}.]{$\Sigma^{*-}$ scaled energy distribution: comparison among SHM predictions, \mbox{\texttt{Herwig6510}} predictions and OPAL data \cite{OPAL4}.}
\label{fig:x_e_sigmastm}
\end{figure}

\begin{figure}
\includegraphics[width=0.450\textwidth]{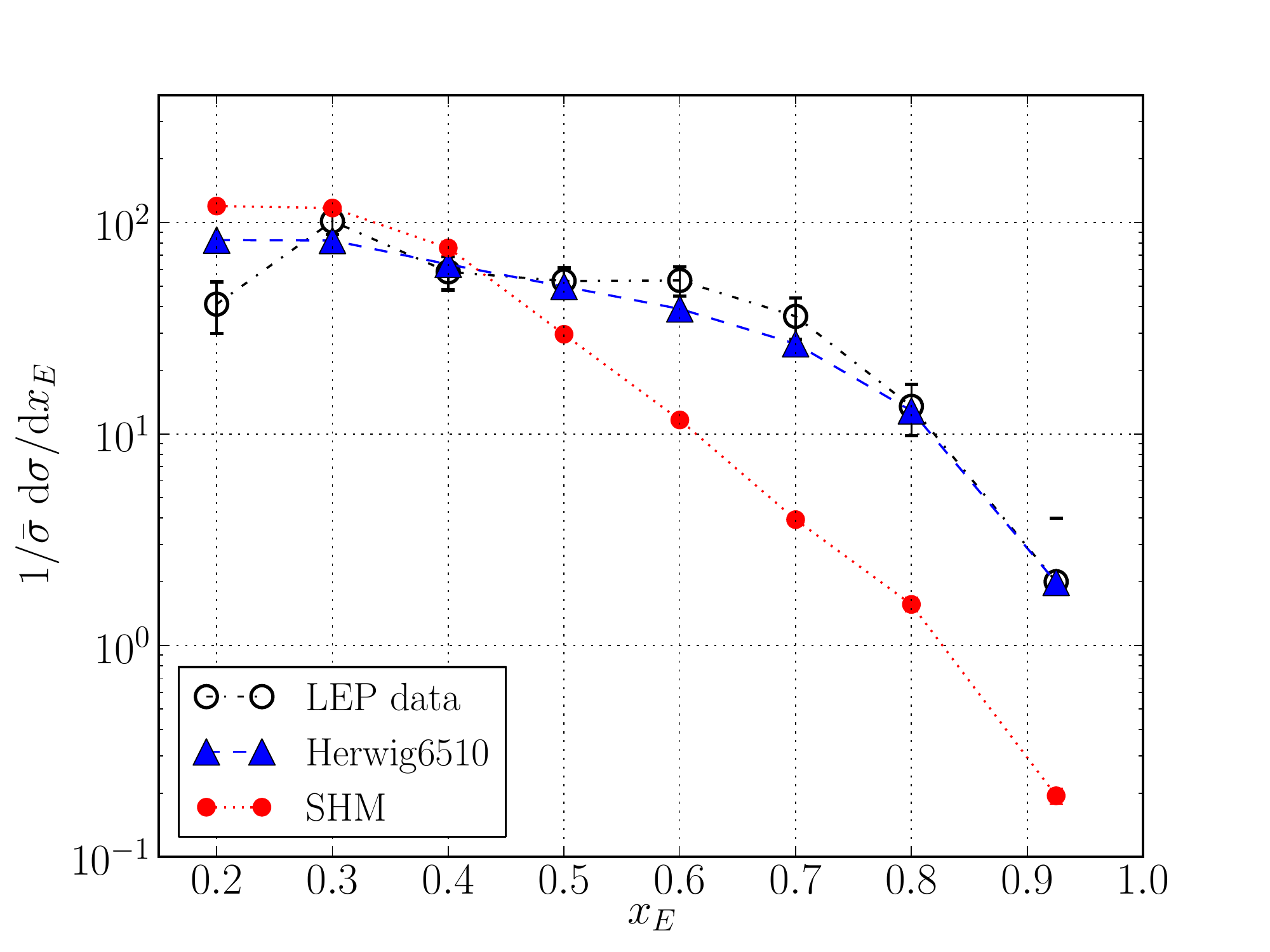}
\caption[$D^{0}$ scaled energy distribution: comparison among SHM predictions, \mbox{\texttt{Herwig6510}} predictions and DELPHI data \cite{DELPHI4}.]{$D^{0}$ scaled energy distribution: comparison among SHM predictions, \mbox{\texttt{Herwig6510}} predictions and DELPHI data \cite{DELPHI4}.}
\label{fig:x_e_Dzero}
\end{figure}

\begin{figure}
\includegraphics[width=0.450\textwidth]{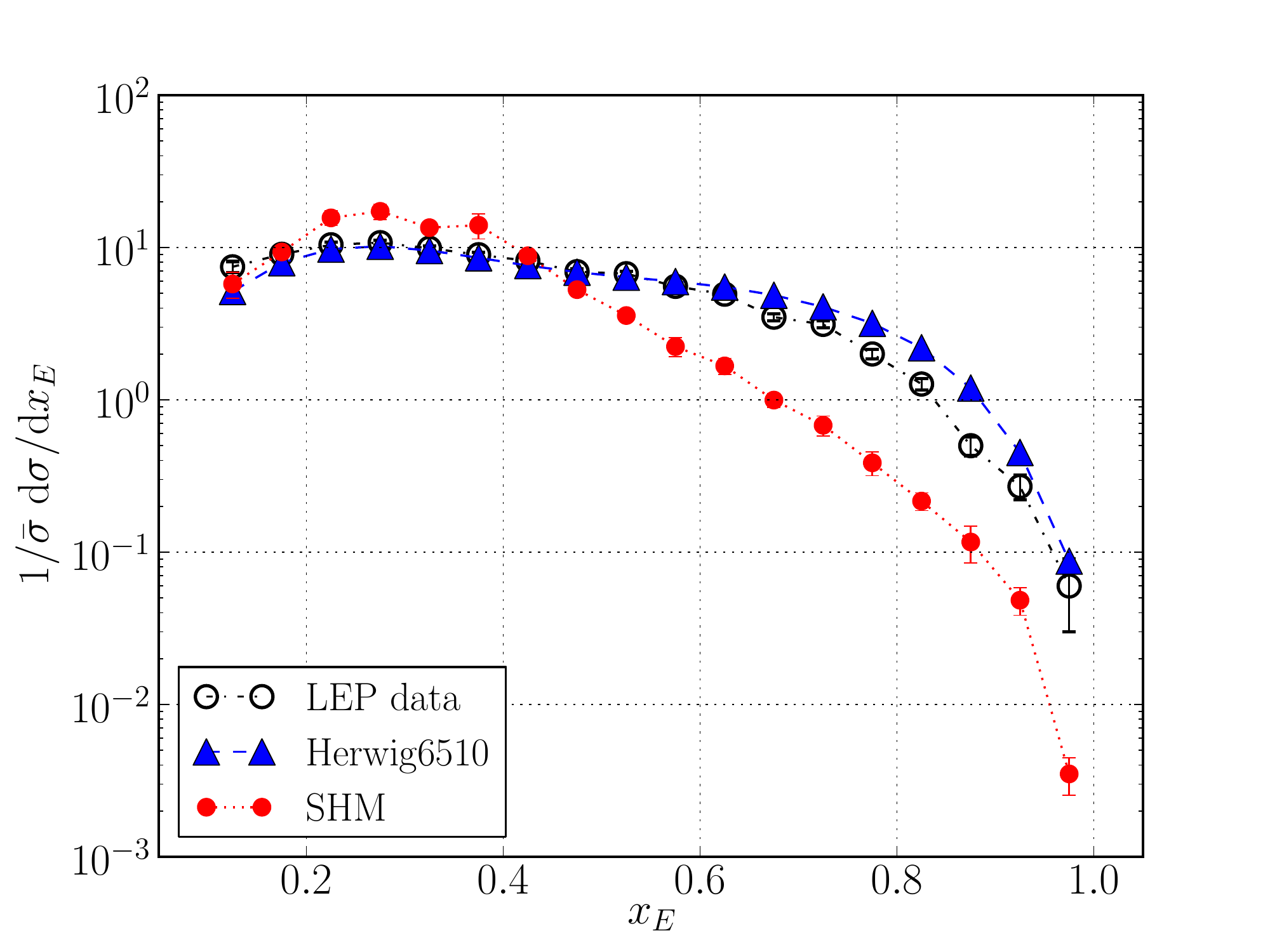}
\caption[$D^{*}$ scaled energy distribution: comparison among SHM predictions, \mbox{\texttt{Herwig6510}} predictions and ALEPH data \cite{ALEPH}.]{$D^{*}$ scaled energy distribution: comparison among SHM predictions, \mbox{\texttt{Herwig6510}} predictions and ALEPH data \cite{ALEPH}.}
\label{fig:x_e_Dstar}
\end{figure}

\begin{figure}[!h]
\includegraphics[width=0.450\textwidth]{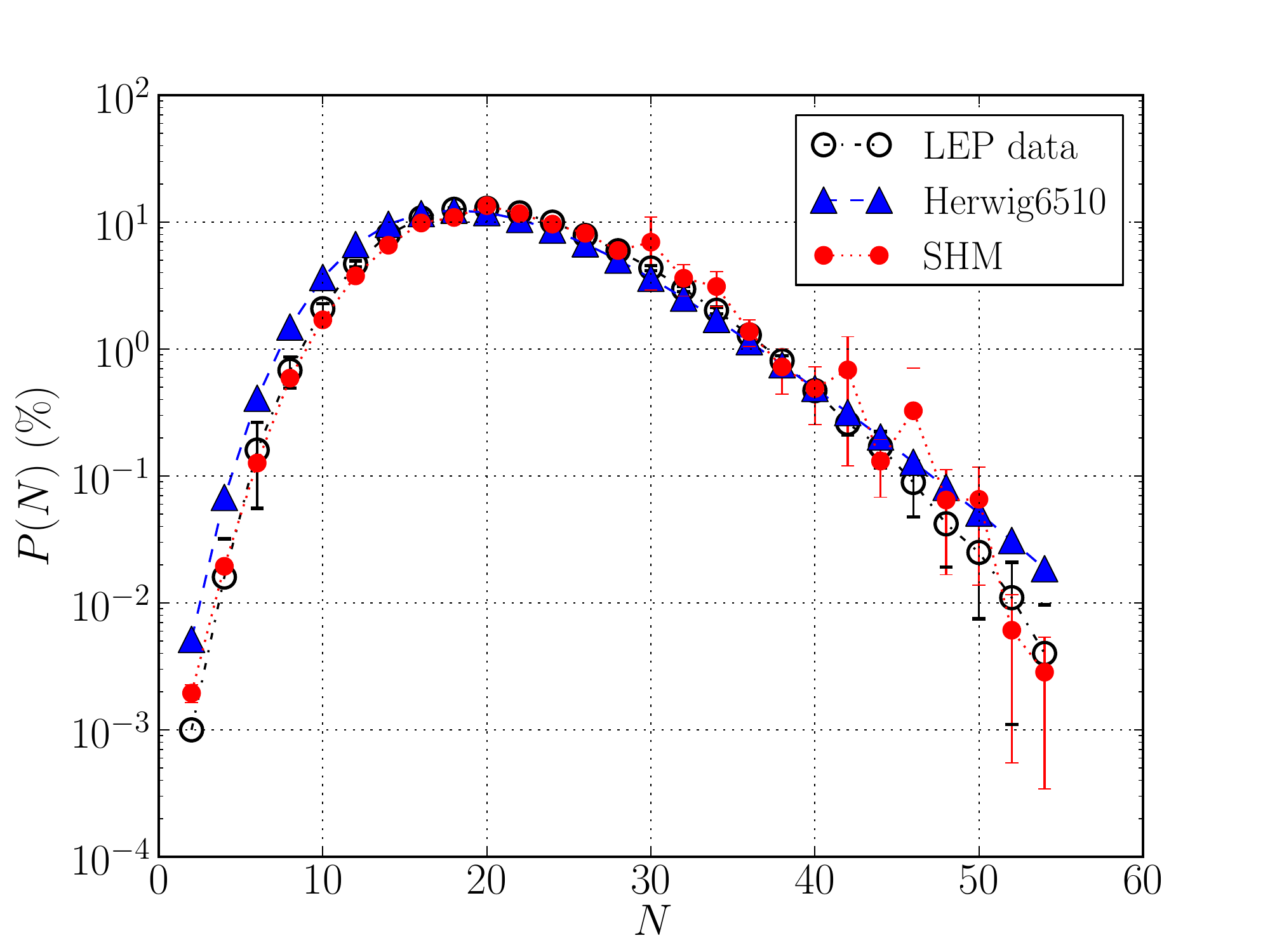}
\caption[Charged particle number probability distribution: comparison among SHM predictions, \mbox{\texttt{Herwig6510}} predictions and OPAL data \cite{OPAL9}.]{Charged particle number probability distribution: comparison among SHM predictions, \mbox{\texttt{Herwig6510}} predictions and OPAL data \cite{OPAL9}.}
\label{fig:Chdmult}
\end{figure}

\begin{itemize}
\item Event shape observables: the comparison between the SHM predictions and LEP 
experimental data shows a quite good global agreement for the whole set of considered observables 
(Figs. \ref{fig:ptT_in} - \ref{fig:1_T}). However the SHM, for the present parameter 
configuration, seems to fail in reproducing the transverse momentum distribution tail behavior of the 
experimental data (Figs. \ref{fig:ptT_in} - \ref{fig:ptT_out}), which is instead correctly predicted by 
\texttt{Herwig6510}. 

\vspace{0.2cm}

\item Single particle observables\footnote{The theoretical distributions have been normalized to 
the integral value of the experimental ones.} (Figs. \ref{fig:p_picharged} - \ref{fig:x_e_Dstar}): also 
in this case a good agreement between the SHM predictions and LEP data is present for a large 
set of the considered single particle observables. Nevertheless, a wrong behavior in the SHM 
predictions can be seen in the $D^{0}$ and $D^{*}$ scaled energy distributions 
(Figs. \ref{fig:x_e_Dzero} - \ref{fig:x_e_Dstar}): in these cases also \texttt{Herwig6510}'s predictions 
show a disagreement with the experimental data, however the Cluster Model seems to be able 
to reproduce the general behavior of these data better than the SHM. A more detailed analysis 
regarding these last results is reported hereafter.

\vspace{0.2cm}

\item Charmed hadrons: the performed analysis, involving a complete collision simulation, 
has given the possibility to identify some possible limits of the SHM, not observed in previous studies. 
In particular, this is the case of the $D^{0}$ and $D^{*}$ scaled energy distributions of Figs. \ref{fig:x_e_Dzero} 
and \ref{fig:x_e_Dstar} respectively, where the SHM fails in reproducing the experimental distribution shapes. 
In these distributions two distinguishable regions are present: a low energy region ($x_{E}\lesssim 0.5$), 
whose filling events correspond to the $D^{0}$ and $D^{*}$ meson production in b-hadron decay, and a 
high energy region approximatively corresponding to the scaled energy distribution of the primary $D^{0}$ and $D^{*}$ 
mesons. While the distribution behavior in these two regions is qualitatively well reproduced by the Cluster Model 
predictions, the SHM in the adopted simulation framework fails in reproducing the high energy region. 
Even though this anomalous result needs to be further investigated, it is understood that the underestimation 
of the charmed mesons scaled energy, provided by the SHM, is strictly related to the large mean number 
of particles produced during the microcanonical hadronization of charmed clusters and to the 
low energy availability condition which follows. A condition not present in the two body cluster 
decays performed by the \texttt{Herwig6510} Cluster Model.

\vspace{0.2cm}

\item Charged particle multiplicity: in this case (Fig. \ref{fig:Chdmult}) a very good agreement between 
the theoretical predictions of the SHM and the experimental data is present. It is worth noting that 
for this distribution, in particular for the distribution tails, the predictions of the SHM show a better 
agreement with the experimental data with respect to \texttt{Herwig6510}'s results.

\vspace{0.2cm}

\item Particle multiplicities: the comparison on the mean particle multiplicities 
(Tabs. \ref{tab:mult2} - \ref{tab:mult4}), for this preli\-minary analysis, shows a global agreement of the 
microcanonical predictions with respect to the expe\-rimental data within $3\sigma$ in the 75.5$\%$ 
of the consi\-dered cases. This percentage becomes $100\%$ if only charmed and bottomed 
hadrons are considered, for which a discrepancy within $1\sigma$ is present in the $55.6\%$ 
of the cases. The corresponding \texttt{Herwig6510} predictions agree with experimental data within 
$3\sigma$ in the $55.1\%$ of cases for the whole set of particles and in the $72.2\%$ of cases if only 
heavy flavored hadrons are considered. More in detail, a different behavior between the SHM and the 
Cluster Model predictions can be observed for the light flavored baryon multiplicities, with the SHM 
tending to underestimate the baryon yields and the \texttt{Herwig6510} hadronization 
model showing the opposite behavior. Finally, a 5$\sigma$ discrepancy between the SHM prediction 
of charged particle total number and the data can be noted. A better agreement for this quantity can 
be obtained with a more refined tuning of the parameters, probably at the cost of a worsening of the 
agreement between data and predictions of other observables (e.g. event shape distributions and 
single particle multiplicities). This discrepancy will be further investigated in the next SHM parameter tuning.

\end{itemize}

\section{Conclusions and outlook}
\label{sec:Concl}

We have presented a Monte-Carlo implementation of Statistical Hadronization Model based on
the microcanonical hadronization of massive clusters. The developed hadronization code has been 
interfaced to the \texttt{Herwig} event generator providing the massive clusters to be 
hadronized. With this method, a full description of an \ee col\-lision at high energy can
be obtained with statistical model based hadronization. 

The results obtained in \ee collisions at the Z peak have been presented and discussed. 
While a final and rigorous tuning of the hadronization module free parameters has not been 
performed in this work, the preliminary comparison confirms the good agreement of the SHM
predictions with the data, already observed in previous studies. Specifically, the agreement
between measured and predicted abundances of hadronic species is confirmed, particularly
in the heavy flavor sector. A good agreement, between SHM predictions and experimental data, 
is also found for a set of hadronization relevant event shape variables and single particle 
energy/momentum distributions, with the only exception of the charmed meson scaled energy, 
where a consistent discrepancy is observed. 
Indeed, the SHM seems to fail in correctly predicting the momentum spectrum at high momentum. 
While this anomaly needs to be further investigated, it is already clear how its origin 
lies in the larger mean number of particles produced by statistical microcanonical hadronization 
of charmed clusters as compared with standard \texttt{Herwig} procedure. 

These problems will be further investigated, and a global fine tuning of the free parameters
performed. We envisage an extension of this hadronization Monte-Carlo code to
high energy ${\rm p}{\rm p}$ and ${\rm p}{\bar{\rm p}}$ collisions, with the final goal of achieving a global assessment 
of the Statistical Hadronization Model in elementary high energy collisions. A first public release
of this code with the needed interfaces for its usage with \texttt{Herwig6510} is forthcoming. 

\begin{acknowledgements}
Work supported in part by the EU Marie Curie Research Training Network MCnet (contract number MRTN-CT-2006-035606). 
C. Bignamini is grateful to S. Gieseke for support and useful discussions and the ITP of the Karslruhe Institute of
Technology for warm hospitality. Moreover, the authors are grateful to L. Ferroni for useful discussions and to 
C. Carloni Calame and G. Zanderighi for providing part of the event analysis routines.  
\end{acknowledgements}

\newpage

\clearpage

\clearpage

\clearpage

\begin{table*}
\begin{center}
\caption[Charged particles, photon and hadron multiplicity mean values for \mbox{${\rm e}^+{\rm e}^- \rightarrow {\rm q} \bar{{\rm q}}$} collisions, with ${\rm q} = {\rm u},{\rm d},{\rm s},{\rm c},{\rm b}$: comparison among SHM predictions, \mbox{\texttt{Herwig6510}} predictions and LEP data \cite{PDG,BecMann} at $91.2$ GeV center of mass energy. The last two columns contain the discrepancies, measured in standard deviations, for SHM and \texttt{Herwig6510} predictions with respect to experimental data.]{Charged particles, photon and hadron multiplicity mean values for \mbox{${\rm e}^+{\rm e}^-\rightarrow {\rm q} \bar{{\rm q}}$} collisions, with ${\rm q} = {\rm u},{\rm d},{\rm s},{\rm c},{\rm b}$: comparison among SHM predictions, \mbox{\texttt{Herwig6510}} predictions and LEP data \cite{PDG,BecMann} at $91.2$ GeV center of mass energy. The last two columns contain the discrepancies, measured in standard deviations, for SHM and \texttt{Herwig6510} predictions with respect to experimental data.}
\label{tab:mult2}
\begin{tabular}{c c c c c c}
\hline\noalign{\smallskip}
 & SHM & \texttt{Herwig6510} & LEP data & $\Delta_{\textnormal{SHM}}$ & $\Delta_{\textnormal{\texttt{Herwig6510}}}$ \\ 
\noalign{\smallskip}\hline\noalign{\smallskip}
Charged &  $ 22.34 \pm 0.27 $ & $20.45$ & $ 20.76 \pm 0.16 $ & $5.00$ & $-1.94$ \\ 
$\gamma$ &  $ 23.64 \pm 0.33 $ & $20.11$ & $ 20.97 \pm 1.17 $ & $2.20$ & $-0.74$ \\ 
$\pi^{0}$ &  $ 10.98 \pm 0.13 $ & $9.56$ & $ 9.61 \pm 0.29 $ & $4.33$ & $-0.19$ \\ 
$\pi^{+}$ & $ 9.33 \pm 0.15 $ & $8.16$ & $ 8.50 \pm 0.10 $ & $4.55$ & $-3.39$ \\ 
$\eta$ & $ 1.18 \pm 0.04 $ & $0.63$ & $  1.059 \pm 0.086  $ & $1.26$ & $-5.05$ \\ 
$\rho^{+}$ & $ 1.16 \pm 0.03 $ & $0.97$ & $  1.20 \pm 0.22 $ & $-0.18$ & $-1.06$ \\ 
$\rho^{0}$ & $ 1.42 \pm 0.04 $ & $1.00$ & $  1.40 \pm 0.13 $ & $0.13$ & $-3.06$ \\ 
$\omega$ & $ 1.29 \pm 0.03 $ & $0.97$ & $ 1.024 \pm 0.059 $ & $4.10$ & $-0.87$ \\ 
$\eta'$ & $ 0.13 \pm 0.01 $ & $0.10$ & $ 0.166 \pm 0.047 $ & $-0.67$ & $-1.40$ \\ 
$f_{0}(980)$ & $0.12 \pm 0.01$ & $0.01$ & $ 0.1555 \pm 0.0085 $ & $-2.43$ & $-17.2$ \\ 
$a_{0}^{+}$ & $ 0.12 \pm 0.01 $ & $0.01$ & $ 0.135 \pm 0.054 $ & $-0.26$ & $-2.30$ \\ 
$\phi$ & $ 0.167 \pm 0.007 $ & $0.1278$ & $ 0.0977 \pm 0.0058 $ & $7.16$ & $5.18$ \\ 
$f_{2}$ & $ 0.17 \pm 0.01 $ & $0.169$ & $ 0.188 \pm 0.020 $ & $-0.74$ & $-0.97$ \\ 
$f_{L1}$ & $ 0.081 \pm 0.005 $ & $0.072$ & $ 0.165 \pm 0.051 $ & $-1.63$ & $-1.82$ \\ 
$f'_{2}$ & $ 0.019 \pm 0.002 $ & $0.012$ & $ 0.0120 \pm 0.0058 $ & $1.19$ & $0.017$ \\ 
$K^{+}$ & $ 1.12 \pm 0.02 $ & $1.05$ & $ 1.127 \pm 0.026 $ & $-0.31$ & $-2.91$ \\ 
$K^{0}$ &  $ 1.07 \pm 0.11 $ & $0.942$ & $ 1.0376 \pm 0.0096 $ & $0.30$ & $-9.97$ \\ 
$K^{*+}$ &  $ 0.34 \pm 0.05 $ & $0.273$ & $ 0.357 \pm 0.022 $ & $-0.37$ & $-3.81$ \\ 
$K^{*0}$ &  $ 0.33 \pm 0.02 $ & $0.274$ & $ 0.370 \pm 0.013 $ & $-1.92$ & $-7.40$\\ 
$K_{2}^{*0}$ &  $ 0.031 \pm 0.004 $ & $0.0361$ & $ 0.036 \pm 0.012 $ & $-0.38$ & $0.006$ \\ 
$p$ &  $ 0.45 \pm 0.02 $ & $0.762$ & $ 0.519 \pm 0.018 $ & $-2.79$ & $13.5$ \\ 
$\Delta^{++}$ &  $ 0.069 \pm 0.003 $ & $0.148$ & $ 0.044 \pm 0.017 $ & $1.47$ & $6.13$ \\ 
$\Lambda$ &  $ 0.128 \pm 0.004 $ & $0.322$ & $ 0.1943 \pm 0.0038 $ & $-11.43$ & $33.6$ \\ 
$\Sigma^{+}$ & $ 0.0253 \pm 0.0009 $ & $0.0667$ & $ 0.0535 \pm 0.0052 $ & $-5.35$ & $2.55$ \\ 
$\Sigma^{-}$ &  $ 0.0233 \pm 0.0008  $ & $0.0548$ & $  0.0410 \pm 0.0037  $ & $-4.68$ & $3.74$ \\ 
$\Sigma^{0}$ & $ 0.034 \pm 0.001 $ & $0.0450$ & $ 0.0389 \pm 0.0041 $ & $-1.23$ & $2.71$ \\ 
$\Sigma^{*+}$ & $ 0.0176 \pm 0.0006 $ & $0.0551$ & $ 0.0118 \pm 0.0011 $ & $4.62$ & $39.4$ \\ 
$\Xi^{-}$ & $ (4.0 \pm 0.1)\times10^{-3} $ & $0.0391$ & $ 0.01319 \pm 0.00050 $ & $-17.7$ & $51.8$ \\ 
$\Xi^{*0}$ & $ (2.12 \pm 0.07)\times10^{-3} $ & $1.84\times10^{-2}$ & $ (2.89 \pm 0.50)\times10^{-3} $ & $-1.53$ & $30.9$ \\ 
$\Omega$ & $ (1.09 \pm 0.04)\times10^{-4} $ & $4.94\times10^{-3}$ & $  (6.2 \pm 1.0)\times10^{-4}  $ & $-5.11$ & $43.2$ \\ 
$n$ & $ 0.51 \pm 0.02 $ & $0.683$ & $ 0.991 \pm 0.054 $ & $-8.49$ & $-5.70$ \\ 
\noalign{\smallskip}\hline
\end{tabular}
\end{center}
\end{table*}

\begin{table*}
\begin{center}
\caption[Mean values of charmed hadron multiplicities for \mbox{${\rm e}^+{\rm e}^- \rightarrow {\rm c} \bar{{\rm c}}$} collisions: comparison among SHM predictions, \mbox{\texttt{Herwig6510}} predictions and LEP data \cite{BecMann} at $91.2$ GeV center of mass energy. The last two columns contain the discrepancies, measured in standard deviations, for SHM and \texttt{Herwig6510} predictions with respect to experimental data.]{Mean values of charmed hadron multiplicities for \mbox{${\rm e}^+{\rm e}^- \rightarrow {\rm c} \bar{{\rm c}}$} collisions: comparison among SHM predictions, \mbox{\texttt{Herwig6510}} predictions and LEP data \cite{BecMann} at $91.2$ GeV center of mass energy. The last two columns contain the discrepancies, measured in standard deviations, for SHM and \texttt{Herwig6510} predictions with respect to experimental data.}
\label{tab:mult3}
\begin{tabular}{c c c c c c}
\hline\noalign{\smallskip}
 & SHM & \texttt{Herwig6510} & LEP data & $\Delta_{\textnormal{SHM}}$ & $\Delta_{\textnormal{\texttt{Herwig6510}}}$ \\ 
\noalign{\smallskip}\hline\noalign{\smallskip}
$D^{+}$ & $ 0.22 \pm 0.02 $ & $0.287$ & $  0.238 \pm 0.024  $ & $-0.53$ & $2.03$ \\ 
$D^{0}$ &  $ 0.54 \pm 0.04$ & $0.577$ & $ 0.559 \pm 0.022 $ & $-0.33$ & $0.82$ \\ 
$D_{\textnormal{s}}$ &  $ 0.110 \pm 0.009 $ & $0.112$ & $ 0.116 \pm 0.036 $ & $-0.18$ & $-0.12$ \\ 
$D^{*+}$ & $ 0.19 \pm 0.02 $ & $0.207$ & $ 0.2377 \pm 0.0098 $ & $-2.40$ & $-3.11$ \\ 
$D^{*0}$ &  $ 0.23 \pm 0.02 $ & $0.210$ & $ 0.218 \pm 0.071$ & $0.11$ & $-0.11$ \\ 
$D^{0}_{1}$ & $ 0.015 \pm 0.001 $ & $0.022$ & $  0.0173 \pm 0.0039 $ & $-0.46$ & $1.30$ \\ 
$D^{*0}_{2}$ & $ 0.033 \pm 0.003$ & $0.030$ & $ 0.0484 \pm 0.008 $ & $-1.85$ & $-2.27$ \\ 
$D^{*}_{\textnormal{s}}$ & $ 0.072 \pm 0.006 $ & $0.036$ & $ 0.069 \pm 0.026 $ & $0.12$ & $-1.26$ \\ 
$D_{\textnormal{s}1}$ & $ 0.0053 \pm 0.0004$ & $0.0044$ & $ 0.0106 \pm 0.0025 $ & $-2.11$ & $-2.50$ \\ 
$D^{*}_{\textnormal{s}2}$ &  $ 0.0038 \pm 0.0003 $ & $0.0059$ & $ 0.0140 \pm 0.0062 $ & $-1.65$ & $-1.31$ \\ 
$\Lambda_{\textnormal{c}}$ & $ 0.13 \pm 0.01$ & $0.036$ & $ 0.079 \pm0.022 $ & $2.19$ & $-1.94$ \\ 
\noalign{\smallskip}\hline
\end{tabular}
\end{center}
\end{table*}

\begin{table*}
\begin{center}
\caption[Mean values of bottomed hadron multiplicities for \mbox{${\rm e}^+{\rm e}^- \rightarrow {\rm b} \bar{{\rm b}}$} collisions: comparison among SHM predictions, \mbox{\texttt{Herwig6510}} predictions and LEP data \cite{BecMann} at $91.2$ GeV center of mass energy. The last two columns contain the discrepancies, measured in standard deviations, for SHM and \texttt{Herwig6510} predictions with respect to experimental data.]{Mean values of bottomed hadron multiplicities for \mbox{${\rm e}^+{\rm e}^- \rightarrow {\rm b} \bar{{\rm b}}$} collisions: comparison among SHM predictions, \mbox{\texttt{Herwig6510}} predictions and LEP data \cite{BecMann} at $91.2$ GeV center of mass energy. The last two columns contain the discrepancies, measured in standard deviations, for SHM and \texttt{Herwig6510} predictions with respect to experimental data.}
\label{tab:mult4}
\begin{tabular}{c c c c c c}
\hline\noalign{\smallskip}
 & SHM & \texttt{Herwig6510} & LEP data & $\Delta_{\textnormal{SHM}}$ & $\Delta_{\textnormal{\texttt{Herwig6510}}}$ \\ 
\noalign{\smallskip}\hline\noalign{\smallskip}
$(B^{0}+B^{+})/2$ & $ 0.411 \pm 0.005 $ & $0.4471$ & $ 0.399 \pm  0.011 $ & $1.03$ & $4.37$ \\ 
$B_{\textnormal{s}}$ & $  0.105 \pm 0.001 $ & $0.108$ & $ 0.098 \pm 0.012 $ & $0.61$ & $0.82$ \\ 
$B^{*}/B_{\textnormal{uds}}$ & $ 0.69 \pm 0.01 $ & $0.424$ & $ 0.749 \pm 0.040 $ & $-1.53$ & $-8.13$ \\ 
$B^{**}$ & $ 0.183 \pm 0.002 $ & $0.143$ & $ 0.180 \pm 0.025 $ & $0.11$ & $-1.48$ \\ 
$(B^{*}_{2}+B_{1})$ & $ 0.121 \pm 0.001 $ & $0.094$ & $ 0.09 \pm 0.018 $ & $1.73$ & $0.19$ \\ 
$B^{*}_{\textnormal{s}2}$ & $ 0.00776 \pm 0.00009 $ & $2\times10^{-8}$ & $ 0.0093 \pm 0.0024 $ & $-0.64$ & $-3.88$ \\ 
$\textnormal{b-baryon}$ & $ 0.110 \pm 0.001 $ & $0$ & $ 0.103 \pm 0.018 $ & $0.37$ & $-5.72$ \\ 
\noalign{\smallskip}\hline
\end{tabular}
\end{center}
\end{table*}

\newpage

\section*{Appendix: observable definitions}
\label{sec:App}

\begin{itemize}

\item The thrust $T$ is defined as: 
\begin{equation*}
T = \max_{\mathbf{n}}\left({\displaystyle\sum_{i=1}^{N} \mid \mathbf{p}_{i}\cdot\mathbf{n} \mid}/{\displaystyle\sum_{i=1}^{N} \mid \vec{p}_{i}\mid}\right),
\end{equation*}
where $\vec{n}$ is a unit vector along the thrust axis, $N$ is the number of particles and $\mathbf{p}_{i}$ 
is the $i$-th particle \mbox{3-momentum}. Thrust major $M$ and minor $m$ are similarly defined, replacing 
$\mathbf{n}$ with $\mathbf{n}_{M}$ (perpendicular to $\mathbf{n}$) and with $\mathbf{n}_{m} = \mathbf{n}_{M}\times\mathbf{n}$ 
respectively.
\item The rapidity with respect to the thrust axis $y_{T}$ is defined as:
\begin{equation*}
y_{T} = \frac{1}{2}\log\left(\frac{E+p_{T}}{E-p_{T}}\right),
\end{equation*}
where $E$ and $p_{T}$ are the particle energy and 3-mo\-mentum projection along the trust axis 
respectively.
\item The in and out components of the transverse momentum with respect to thrust axis, $p^{in}_{T}$ and 
$p^{out}_{T}$,  are defined as: 
\begin{equation*}
\begin{split}
& p^{in}_{T} = \mathbf{p}\cdot\mathbf{n}_{M}\\
& p^{out}_{T} = \mathbf{p}\cdot\mathbf{n}_{m},
\end{split}
\end{equation*}
where $\mathbf{p}$ is the particle 3-momentum and $\mathbf{n}_{M}$ ($\mathbf{n}_{m}$) is the thrust major (minor) 
axis previously defined. 
\item The scaled energy and momentum $x_{E}$, $x_{p}$ and $\xi_{p}$ are defined as: 
\begin{equation*}
\begin{split}
& x_{E} = \frac{2E}{\sqrt{s}}\\
& x_{p} = \frac{2|\mathbf{p}|}{\sqrt{s}}\\
& \xi_{p} = -\log\left(x_{p}\right),
\end{split}
\end{equation*}
where $E$ and $\mathbf{p}$ are the particle energy and 3-momen\-tum respectively and $\sqrt{s}$ is the collision 
center of mass energy.
\end{itemize}

\end{document}